\documentclass[12pt]{article}
\usepackage{epsfig}
        \newdimen\eqskip
        \newdimen\txtskip
        \baselineskip=\txtskip
        \newdimen\mysep                
        \newdimen\hmysep
\begin{document}
\renewcommand{\thefootnote}{\fnsymbol{footnote}}
  \newcommand{\ccaption}[2]{
    \begin{center}
    \parbox{0.85\textwidth}{
      \caption[#1]{\small{{#2}}}
      }
    \end{center}
    }
\newcommand{\BS}{\bigskip}
% MATH SYMBOLS
\def    \be             {\begin{equation}}
\def    \ee             {\end{equation}}
\def    \beq             {\begin{equation}}
\def    \eeq             {\end{equation}}
\def    \ba             {\begin{eqnarray}}
\def    \ea             {\end{eqnarray}}
\def    \beqn           {\begin{eqnarray}}
\def    \eeqn           {\end{eqnarray}}
\def    \beeq           {\begin{eqnarray}}
\def    \eeeq           {\end{eqnarray}}
\def    \nn             {\nonumber}
\def    \=              {\;=\;}
\def    \frac           #1#2{{#1 \over #2}}
\def    \ret            {\\[\eqskip]}
\def    \ie             {{\em i.e.\/} }
\def    \eg             {{\em e.g.\/} }
\def \lsim{\mathrel{\vcenter
     {\hbox{$<$}\nointerlineskip\hbox{$\sim$}}}}
\def \gsim{\mathrel{\vcenter
     {\hbox{$>$}\nointerlineskip\hbox{$\sim$}}}}
\def    \bentarrow      {\:\raisebox{1.1ex}{\rlap{$\vert$}}\!\rightarrow}
\def    \rd             {{\mathrm d}}    
\def    \Im             {{\mathrm{Im}}}  
\def    \bra#1          {\mbox{$\langle #1 |$}}
\def    \ket#1          {\mbox{$| #1 \rangle$}}
\def    \to             {\rightarrow} 

% UNITS                 
\def    \kev            {\mbox{$\mathrm{keV}$}}
\def    \mev            {\mbox{$\mathrm{MeV}$}}
\def    \gev            {\mbox{$\mathrm{GeV}$}}

% KINEMATICAL VARIABLES 

\def    \mq             {\mbox{$m_Q$}}  
\def    \mt             {\mbox{$m_t$}}  
\def    \mb             {\mbox{$m_b$}}  
\def    \mqq            {\mbox{$m_{Q\bar Q}$}}
\def    \mqqsq          {\mbox{$m^2_{Q\bar Q}$}}
\def    \pt             {\mbox{$p_T$}}
\def    \ptsq           {\mbox{$p^2_T$}}

% QCD PARAMETERS                                      
\newcommand     \MSB            {\ifmmode {\overline{\rm MS}} \else 
                                 $\overline{\rm MS}$  \fi}
\def    \muf            {\mbox{$\mu_{\rm F}$}}
\def    \mug            {\mbox{$\mu_\gamma$}}
\def    \mufsq          {\mbox{$\mu^2_{\rm F}$}}
\def    \mur            {{\mbox{$\mu_{\rm R}$}}}
\def    \mursq          {\mbox{$\mu^2_{\rm R}$}}
\def    \mul            {{\mu_\Lambda}}
\def    \mulsq          {\mbox{$\mu^2_\Lambda$}}

\def    \bzero          {\mbox{$b_0$}}
\def    \as             {\ifmmode \alpha_s \else $\alpha_s$ \fi}
\def    \asb            {\mbox{$\alpha_s^{(b)}$}}
\def    \assq           {\mbox{$\alpha_s^2$}}
\def \oacube {\mbox{$ O(\alpha_s^3)$}}
\def \oafour {\mbox{$ O(\alpha_s^4)$}}
\def \oatwo {\mbox{$ O(\alpha_s^2)$}}
\def \oas   {\mbox{$ O(\alpha_s)$}}
\def\asp{{\alpha_s}\over{\pi}}

\def\slash#1{{#1\!\!\!/}}
\def\rt1{\raisebox{-1ex}{\rlap{$\; \rho \to 1 \;\;$}}
\raisebox{.4ex}{$\;\; \;\;\simeq \;\;\;\;$}}
\def\ltap{\raisebox{-.5ex}{\rlap{$\,\sim\,$}} \raisebox{.5ex}{$\,<\,$}}
\def\gtap{\raisebox{-.5ex}{\rlap{$\,\sim\,$}} \raisebox{.5ex}{$\,>\,$}} 

\def \ee  {e^+e^-}
\def\GE{\gamma_E}
\def\half{\frac{1}{2}}
\def\b0{\beta_0}
\def\naive{na\"{\i}ve}
\def\cm{{\cal M}}
\def\bom#1{\mbox{\bf{#1}}}
%%%%%%%%%%%%%%%%%%%%%%%%%%%%%%%%%%%%%%%%%%%%%%%%%%%%%%%%%%%%%%%%%%%%%%
\begin{titlepage}
\nopagebreak       {\flushright{
        \begin{minipage}{5cm}
        CERN-TH/98-214\\
        {\tt hep-ph/9806484}\\
        \today \\
        \end{minipage}        }

}
\vfill
\begin{center}
{\LARGE { \bf \sc Sudakov Resummation for Prompt-Photon \\[0.5cm]
Production in Hadron Collisions~\footnote{This work was supported in part 
by the EU Fourth Framework Programme ``Training and Mobility of Researchers'', 
Network ``Quantum Chromodynamics and the Deep Structure of
Elementary Particles'', contract FMRX--CT98--0194 (DG 12 -- MIHT).}}}
\vfill                                                       
{\bf      Stefano CATANI
     \footnote{On leave of absence from INFN, Sezione di Firenze, Italy.},
          Michelangelo L. MANGANO                                
    \footnote{On leave of absence from INFN, Sezione di Pisa, Italy.}}
     and 
{\bf     Paolo NASON
    \footnote{On leave of absence from INFN, Sezione di Milano, Italy.}
}\\[1cm]

{CERN, Theoretical Physics Division, \\ CH~1211 Geneva 23, Switzerland} 
\end{center}                             
\nopagebreak
\vfill
%\vskip 3cm
\begin{abstract} 
We present the explicit expressions for the resummation of
large-$x_T$ Sudakov effects in the transverse-energy distribution of 
prompt-photons produced in hadronic collisions,      
to next-to-leading logarithmic (NLL) accuracy. Fragmentation processes do
not contribute to the Sudakov resummation at NLL level.
In Mellin space, the resummed radiative factor
factorizes in the product of independent radiators for the two initial and the
one final coloured partons appearing in the Born process, times a simple factor
describing the soft-gluon interferences between initial and final states.  The
formulae are given in terms of Mellin moments, and can be used for 
phenomenological applications using standard techniques  for the inverse-Mellin
transforms. The calculations presented in this work, when added to the existing
works on DY and DIS production, complete the theoretical       
ground-work necessary to carry out global fits of parton densities with a
uniform  NLL accuracy in the large-$x$ region. 
\end{abstract}                                                
\vskip 1cm
CERN-TH/98-214\hfill \\
June 1998     
\vfill       
\end{titlepage}

\section{Introduction}
\label{sec:1}\label{introduction}
Prompt-photon production  plays a very important role in our understanding of
the physics of hadron collisions. At the leading order (LO) in QCD perturbation
theory, prompt photons are produced via light-quark annihilation, with emission
of a hard gluon recoiling against the photon, or via quark-gluon Compton
scattering, with the emission of a quark. When the Compton process dominates
the cross section, tests or even measurements of the gluon density inside the
proton can be performed. This is the case of photon production at small 
$x_T= 2E_T/{\sqrt S}$ ($x_T \lsim 0.1$) in $p\bar p$ collisions, 
and it is true for all values of $x_T$ that are accessible
in fixed-target proton-nucleon collisions, due to smaller content of
antiquarks relative to gluons in the nucleon sea. In particular, prompt-photon
production at large $x_T$ can therefore be used                        
to constrain or measure the gluon density at large $x$.
This region of the gluon density is of great importance for the
study of high-transverse-momentum phenomena at hadron colliders,
and it is not accessible using only Deep-Inelastic-Scattering (DIS) data.

The cross section for inclusive photon production has been computed at the
next-to-leading order (NLO) in perturbation theory 
\cite{aurenche}--\cite{gordon93}.
%\cite{aurenche, baer90, gordon93}. 
The NLO computation of {\em isolated} photon
production~\cite{berger91}--\cite{pilon},
%~\cite{berger91,baer90}, 
which is the relevant quantity for the
measurements carried out in high-energy $p\bar p$ collisions, 
%has also been completed~\cite{gordon94}. 
is also available in the limit of small size of the isolation 
cone~\cite{bailey92}--\cite{gordon94}.
These calculations include all
light-parton  fragmentation processes up to NLO~\cite{ACGG}. 
                                                                          
These theoretical results have been used~\cite{aurenche89}--\cite{vogelsang95}
to probe the overall consistency of the
prompt-photon production data from both
fixed-target~\cite{WA70}--\cite{E706-93} and 
collider experiments~\cite{ISR}--\cite{CDF-D0}. 
                              
The interpretation of the data has not provided so far a fully satisfactory
picture. The study by Huston et al.~\cite{huston95} exposed a tendency of the
$x_T$ distributions to be steeper than theory, regardless of the value of
$\sqrt{S}$. This result could not be accomodated by a simple modification of
the parton densities, since different experiments probe different values of
$x_T$. These authors therefore proposed that additional mechanisms should be
introduced to explain the pattern of the data. In the fixed-target regime, such
a mechanism would be provided by the
presence of a non-perturbative $k_T$ kick, which would smear the $E_T$ spectra.
This phenomenon was also advocated to help explain the spectra of
fixed-target heavy-quark production~\cite{fmnr}. In the high-energy
regime, probed by the Tevatron experiments, the $E_T$ smearing necessary to
reconcile theory and data could be                      
%is naturally
provided by the inclusion of multi-gluon emission effects from the evolution of
the initial state, as advocated by Baer and Reno~\cite{baer96}. 
                                                       
%A further 
The analysis by Vogelsang and Vogt~\cite{vogelsang95} indicated that
allowing for different choices of  factorization and renormalization scales, a
satisfactory fit to the data could be accomodated by modifying the
gluon density within the range allowed by the DIS data available in 1995.  This
interpretation is apparently not viable anymore~\cite{vogelsang98}, because of
the most recent constraints on the gluon density extracted at small $x$ from 
the HERA data, and, in particular, because of
the latest fixed-target prompt-photon data from E706~\cite{E706-97}.
                         
The comparison of the E706 data with NLO QCD, carried out in~\cite{E706-97},
seems to confirm the need for an intrinsic-$k_T$ smearing corresponding to
$\langle k_T \rangle \sim 1$~GeV.  These conclusions are shared in a recent
global fit of the parton densities performed by the MRST group~\cite{MRST}.
In this same study (see also \cite{fontannaz}),
however, a strong dependence of $\langle k_T \rangle$ on
$\sqrt{S}$ is claimed to be necessary to properly describe the lower-energy data
published by WA70. 

In conclusion, the comparison of the large-$x_T$
fixed-target data with NLO QCD still presents some puzzling features, which 
will need to be properly clarified before use of these data can be made to
place robust constraints on the large-$x$ gluon distribution inside the proton. 
This is unfortunate, since these data provide today the only independent probe
on high-$x$ gluons. Their accurate interpretation is therefore a fundamental
ingredient for an accurate prediction of the production rate of
high-$E_T$ jets at the Tevatron, a measurement which has challenged 
perturbative QCD in the recent past~\cite{cdfjets}.

To improve the reliability of the perturbative predictions for the production of
prompt photons at large $x_T$, and detect the presence of potentially large
corrections beyond NLO that could change the interpretation of the current
data, in the present work  we consider                               
an extension of the NLO formalism
that includes large logarithmically-enhanced effects
as the production threshold is approached.
This is the kinematical region of interest for the fixed-target data.
As the $x_T$ of the photon is increased, the parton luminosity becomes
steeper, being driven down by the strong suppression of the gluon
density at large $x$. We thus enter a regime of inhibited radiation:
further radiation of soft gluons is strongly suppressed,
and logarithmically-enhanced effects (Sudakov effects) arise
at any order in the perturbative expansion. These
logarithms spoil the reliability of the fixed-order expansion in the strong
coupling $\as$ and, hence, their summation to all pertubative orders is 
necessary. For simplicity, this regime can be described in terms
of the distance from the kinematic threshold, which is reached            
when $x_T \sim 1$. In
this limit, the coefficients of the perturbative series for the cross
section are enhanced by powers of $\ln (1-x_T)$ that have to be resummed
at all orders.
This simplified description applies to the case of hypothetical structure 
functions
that are not strongly suppressed at high $x$. It is, however,
an appropriate framework for the classification of the perturbative
corrections we are interested in. 
 
In this work we will present all the formalism that is needed to compute
the resummed cross section for direct photons, integrated over the photon 
rapidity and at fixed transverse energy. In particular, we give
explicit resummation formulae that are valid up to next-to-leading
logarithmic (NLL) accuracy.
No phenomenological applications
will be discussed here, but they will be explored in a forthcoming work.
Furthermore, the formulae for the resummed           
correction factors will be presented and illustrated, but not derived
here. The general formalism~\cite{bonciani}
used to obtain the resummation factors, which
has already been used for the NLL resummation of the heavy-quark total
production cross-section in Ref.~\cite{BCMN},
will be presented in a forthcoming publication \cite{inprep}.
                                                           
The rest of this work is organised as follows.
The general theoretical framework is discussed in Sect.~2. In Sect.~3 
we fix our notation
and present the formulae for the Born cross section, together with their
Mellin transforms. Soft-gluon resummation at large $x_T$ is considered
in Sect.~4.
The NLL resummation factors are presented in Sects.~4.1, 4.2. In Sect.~4.3,
the fixed-order expansion of the resummed formulae is 
%at fixed order will be presented, and 
compared with the NLO results of Refs.~\cite{aurenche,gordon93}.
This comparison is also exploited to fix certain
constant factors in the resummed formulae. In Sect.~5 we discuss
similarities and differences between the resummed
factors for the prompt-photon cross section and those for other hard-scattering
processes, and we prove the consistency of the results obtained in the case of
prompt photons with the coherence properties of large-angle soft-gluon
emission. Section~6 contains our conclusions.

More technical details are left to the Appendices.
In Appendix~A, we give the NLL formulae for the radiative factors in the
Mellin transform representation.
Previous experience in the case of heavy-flavour production has
shown that this is what is needed to perform a reliable
phenomenological analysis \cite{BCMN}.
In Appendix~B, the threshold
%large $N$ 
limit of the partonic cross sections
is discussed. In particular, a prediction for the logarithmic terms at
the next-to-next-to-leading order (NNLO) is given.
Finally, in Appendix~C the simpler case of photoproduction of direct photons
is discussed.
 
While completing this paper, a study of the NLL resummation for
single-inclusive distributions, covering the case of prompt-photon production, 
has been released by Laenen, Oderda and Sterman~\cite{Laenen98}.

\section{General framework}
\label{secgen}
The presence of large logarithmically-enhanced contributions is
a common feature in the study of the production cross 
sections of systems of high mass or high transverse energy near threshold.
In this kinematic regime, known as the Sudakov regime, only additional 
soft gluons can be produced.
The radiative tail of the real emission is thus strongly suppressed
and cannot balance the virtual corrections. The imperfect compensation
between real and virtual terms leads to the large logarithmic contributions.

General techniques for resumming soft-gluon corrections 
to hadroproduction processes have been 
developed over the past several years, starting from the case of
Drell-Yan (DY) pair production~\cite{Sterman,CT}.
The resummation program of the soft-gluon contributions is best carried 
out in the Mellin-transform space, or $N$-space, where $N$ denotes the
parameter that is conjugate to the kinematic variable that measures the
distance from threshold. 
In $N$-moment space the threshold-production region corresponds to the limit
$N\to \infty$ and the typical structure of the logarithmic contributions
is as follows  
\beq
\label{genlog}
{\hat \sigma}_N^{(0)} \left\{ 1 + \sum_{n=1}^{\infty} \as^n \sum_{m=1}^{2n}
c_{n,m} \ln^m N \right\} \;\;,
\eeq 
where ${\hat \sigma}_N^{(0)}$ is the corresponding partonic cross-section at
%the leading order 
LO. In the DY
process the logarithmic terms in the curly bracket of Eq.~(\ref{genlog})
can be explicitly summed and organized in a radiative factor $\Delta_{DY,N}$
that has an exponential form~\cite{Sterman,CT,CT2}:
\beqn 
\label{deltady}
\Delta_{DY,N}(\as) &=& \exp \left\{ \sum_{n=1}^{\infty} \as^n 
\sum_{m=1}^{n+1} G_{nm} \ln^m N \right\} \\
\label{deltadyex}
&=& \exp \left\{ \ln N \,g_{DY}^{(1)}(\as \ln N) +  g_{DY}^{(2)}(\as \ln N)
+ \as g_{DY}^{(3)}(\as \ln N) + \dots \right\} \;\;.
\eeqn
Note that the exponentiation in Eq.~(\ref{deltady}) is not trivial. The sum
over $m$ in Eq.~(\ref{genlog}) extends up to $m=2n$ while in
Eq.~(\ref{deltady}) the maximum value for $m$ is smaller, $m \leq n+1$. In
particular, this means that all the double logarithmic (DL) terms
$\as^n c_{n,2n} \ln^{2n} N$ in eq.~(\ref{genlog}) are taken into account
by simply exponentiating the lowest-order contribution 
$\as c_{1,2} \ln^2 N$. Then, the exponentiation in Eq.~(\ref{deltady})
allows one to define the improved perturbative expansion in 
Eq.~(\ref{deltadyex}). The function $\ln N \,g_{DY}^{(1)}$
resums all the {\em leading} logarithmic (LL) contributions 
$\as^n \ln^{n+1} N$, $g_{DY}^{(2)}$ contains the {\em next-to-leading}
logarithmic (NLL) terms $\as^n \ln^n N$,
$\as g_{DY}^{(3)}$ contains the {\em next-to-next-to-leading}
logarithmic (NNLL) terms $\as^{n+1} \ln^n N$,
and so forth. Once the functions
$g_{DY}^{(k)}$ have been computed, one has a systematic perturbative
treatment of the region of $N$ in which $\as \ln N \ltap 1$, which is much
larger than the domain $\as \ln^2 N \ll 1$ in which the fixed-order calculation
in $\as$ is reliable.

The QCD exponentiation formula for the DY process
formally resembles analogous results for QED. This is because the underlying
hard-scattering subprocess involves only {\em two} QCD partons, i.e. the
annihilating $q{\bar q}$ pair, and, hence, both its kinematics and its 
colour structure are simple.
In the case of prompt-photon production, instead, all the LO
hard-scattering subprocesses 
\beq
\label{loproc}
q + {\bar q} \to g + \gamma \;, 
\quad
q + g \to q + \gamma \;,
\quad
{\bar q} + g \to {\bar q} + \gamma \;,
\eeq
involve {\em three} coloured partons and, then, Sudakov resummation is by far
less trivial.

A key ingredient for the exponentiation in the DY process
is the factorization of the corresponding multigluon matrix elements
in the soft limit. Since the colour structure of a two-parton hard-scattering 
is trivial\footnote{This is the reason why similar exponentiation formulae
apply to many other  {\em two-jet}-dominated processes \cite{SC}.},
primary soft radiation from the two hard partons factorizes as 
in QED. Then the subsequent parton radiation can be factorized in 
non-interfering angular-ordered cascades because of the 
coherence properties \cite{BCM} of QCD emission. 

In the case of prompt-photon production, and, in general, in scattering
processes produced by hard interactions of more than two QCD partons,
the colour and momentum flows in the partonic subprocess are more 
involved. In particular, the interplay between colour exchange in the hard
scattering and colour transitions induced by parton radiation spoils
QED-like factorization of soft-gluon emission. Therefore, both
{\em colour correlations} and {\em soft-gluon interferences} 
have to be properly taken into account. It turns out that, in general, the 
threshold logarithmic corrections
cannot be resummed in a single exponential factor \cite{kidon}:
one has to deal with exponential matrices that couple the various colour
channels of the hard-scattering subprocess.

However, the three-parton subprocesses in Eq.~(\ref{loproc}) are a 
special case among the multiparton configurations. There is only 
{\em one colour-singlet} state\footnote{In other words, the $q{\bar q}g$
colour-amplitude ${\cal M}_{q{\bar q}g}^{\alpha {\bar \alpha} a}$ 
($\alpha, {\bar \alpha}$ and $a$ are the colour indices of the quark, 
antiquark and gluon, respectively) is
necessarily proportional to the matrix $t^a_{\alpha {\bar \alpha}}$ of the
fundamental representation of the gauge group $SU(N_c)$.}
that can be constructed by combining 
$q{\bar q}g$ and then, because of colour 
conservation, soft-gluon radiation cannot induce colour transitions in the 
hard-scattering subprocess. Owing to the absence of colour correlations,
we conclude that the logarithmically-enhanced
threshold corrections in prompt-photon hadroproduction are embodied by three
radiative factors
(one factor for each of the LO partonic channels in Eq.~(\ref{loproc})) 
\beq
\label{radfact}
\Delta_N^{q{\bar q} \to g \gamma}(\as) \;, 
\quad
\Delta_N^{qg \to q \gamma}(\as) \;,
\quad
\Delta_N^{{\bar q}g \to {\bar q} \gamma}(\as) \;,
\eeq
that, after all-order resummation, have an exponential form analogous to the
DY radiative factor in Eq.~(\ref{deltadyex}). Nonetheless, the similarity with
the DY process regards only the colour structure. The hard-scattering 
kinematics is different in prompt-photon production and the 
factors in Eq.~(\ref{radfact}) still contain soft-gluon 
{\em interference effects}
that are non-trivial. The pattern of these soft-gluon interferences is typical
of multiparton hard-scatterings.

General theoretical methods to perform Sudakov resummation in
processes initiated by hard scattering of more than two QCD partons 
have recently been developed by two groups. The KOS formalism 
\cite{kidon}--\cite{kos} uses the Wilson line approach
to treat colour correlations and soft-gluon interferences. It
has been explicitly applied to the calculation with NLL accuracy
of the the invariant-mass distributions of heavy-quark pairs and dijets (see
also ref.~\cite{Laenen98}). 
The more recent BCMN formalism \cite{BCMN} is based on generalized 
soft-gluon factorization and has been used for the NLL calculation of
the total cross section for heavy-quark hadroproduction. The consistency
of the NLL results for the total cross section \cite{BCMN} with those
for the invariant mass distribution \cite{kidon} of heavy-quark pairs
shows that, although different, the two formalisms are equivalent to a large
extent. 

In the rest of this paper, we first introduce our notation and then 
we present the resummed expressions of the prompt-photon
radiative factors (\ref{radfact}) to NLL accuracy. The results include
the complete soft-gluon interferences to this accuracy, as
evaluated by using the BCMN formalism. Details of our general
formalism will be presented elsewhere \cite{inprep}.

\section{Notation and fixed-order calculations}
\label{secnotat}

We consider the inclusive production of a single
prompt photon in hadron collisions:
\beq
\label{procgamma}
H_1(P_1) + H_2(P_2) \to \gamma(p) + X \;\;.
\eeq
The colliding hadrons $H_1$ and
$H_2$ respectively carry momenta $P_1^\nu$ and $P_2^\nu$. In their 
centre-of-mass frame, using massless kinematics, they have the following
light-cone coordinates
\beq
\label{mom}
P_1^\nu = {\sqrt \frac{S}{2}} \;(1,{\bom 0},0) \;, \;\;\;\;
P_2^\nu = {\sqrt \frac{S}{2}} \;(0,{\bom 0},1) \;\;,
\eeq
where $S= (P_1+P_2)^2$ is the centre-of-mass energy squared.
The photon momentum $p$ is thus parametrized as
\beq
\label{momp}
p^\nu = \left( \frac{E_T}{{\sqrt 2}} \,e^{\eta}, {\bom E}_T,
\frac{E_T}{{\sqrt 2}} \,e^{- \eta} \right) \;\;,
\eeq
where $E_T$ and $\eta$ are the transverse energy and the pseudorapidity,
respectively. We also introduce
the customary scaling variable $x_T$ $(0 \leq x_T \leq 1)$:
\beq
\label{xt}
x_T = 2 \frac{E_T}{\sqrt S} \;\;.
\eeq

We are interested in the prompt-photon production cross section integrated
over $\eta$  at fixed $E_T$. According to perturbative QCD, the cross section
is given by the following factorization formula
\beeq
\label{1pxsgamma}
\frac{d\sigma_{\gamma}(x_T,E_T)}{d E_T} &=& \frac{1}{E_T^3} \sum_{a,b}
\int_0^1 dx_1 \;f_{a/H_1}(x_1,\mu_F^2)
\,\int_0^1 dx_2 \,\;f_{b/H_2}(x_2,\mu_F^2) \nonumber \\
&\cdot& \int_0^1 dx \left\{
\delta\!\left(x - \frac{x_T}{{\sqrt {x_1 x_2}}} \right) 
{\hat \sigma}_{ab\to {\gamma}}(x, \as(\mu^2); E_T^2, 
\mu^2, \mu_F^2, \mu_f^2) \right. \\
&+& \left. \sum_{c} \int_0^1 dz \;z^2 \;d_{c/\gamma}(z,\mu_f^2)
\;\delta\!\left(x - \frac{x_T}{z{\sqrt {x_1 x_2}}} \right)
\;{\hat \sigma}_{ab\to c}(x, \as(\mu^2); E_T^2, \mu^2, \mu_F^2, \mu_f^2)
\right\} \;\;. \nonumber
\eeeq
where $a,b,c$ denotes the parton indices $(a=q,{\bar q},g)$, and 
$f_{a/H_1}(x_1,\mu_F^2)$ and $f_{b/H_2}(x_1,\mu_F^2)$ are the
parton densities of the colliding hadrons evaluated at the 
factorization scale $\mu_F$.
The first and the second term in the curly bracket on the right-hand side of 
Eq.~(\ref{1pxsgamma})
respectively represent the {\em direct} and the {\em fragmentation} component
of the cross section. The fragmentation component involves the
parton fragmentation function $d_{c/\gamma}(z,\mu_f^2)$
of the observed photon at the factorization scale $\mu_f$, which, in general,
differs from the scale $\mu_F$ of the parton densities.

The {\em rescaled}\footnote{These functions are related to the partonic
differential cross sections by ${\hat \sigma}_{ab\to i} = E_T^3 
d{\hat \sigma}_{ab\to i} / dE_T$ ($i=\gamma,c$).}
partonic cross sections ${\hat \sigma}_{ab\to \gamma}$ 
and ${\hat \sigma}_{ab\to c}$ in Eq.~(\ref{1pxsgamma}) are
computable in QCD perturbation theory as power series expansions in the
running coupling $\as(\mu^2)$, $\mu$ being the renormalization scale
in the $\MSB$ renormalization scheme:
\beq
\label{pxsg}
{\hat \sigma}_{ab\to \gamma}(x, \as(\mu^2); E_T^2, \mu^2, \mu_F^2, \mu_f^2)
= \alpha \, \as(\mu^2) \left[
{\hat \sigma}_{ab\to d \gamma}^{(0)}(x) +
\sum_{n=1}^{\infty} \as^n(\mu^2) \, 
{\hat \sigma}_{ab\to \gamma}^{(n)}(x; E_T^2, \mu^2, \mu_F^2, \mu_f^2)
\right] \;,
\eeq
\beq
\label{pxsc}
{\hat \sigma}_{ab\to c}(x, \as(\mu^2); E_T^2, \mu^2, \mu_F^2, \mu_f^2)
= \as^2(\mu^2) \left[
{\hat \sigma}_{ab\to c}^{(0)}(x) +
\sum_{n=1}^{\infty} \as^n(\mu^2) \, 
{\hat \sigma}_{ab\to c}^{(n)}(x; E_T^2, \mu^2, \mu_F^2, \mu_f^2)
\right] \;.
\eeq
Note that the ratio between the direct
and the fragmentation terms in Eqs.~(\ref{pxsg}) and (\ref{pxsc})
is of the order of $\alpha/\as$, where $\alpha$ is the fine structure constant.
This ratio is compensated by the
photon-fragmentation function $d_{c/\gamma}$, which (at least formally) is of 
the order
of $\alpha/\as$, so that direct and fragmentation components equally
contribute to Eq.~(\ref{1pxsgamma}). 

Throughout the paper we always use parton densities and parton 
fragmentation functions as defined in the $\MSB$ factorization scheme.
In general, we consider different values for the renormalization and 
factorization scales $\mu, \mu_F, \mu_f$, although we always assume that all of
them are of the order of the photon transverse energy $E_T$. 

The LO terms ${\hat \sigma}_{ab\to d \gamma}^{(0)}$ in Eq.~(\ref{pxsg})
are due to the tree-level parton scatterings
\beq
\label{parprogamma}
a + b \to d + \gamma \;\;,
\eeq
where the flavour indices $a,b,d$ are those explicitly denoted in the
subprocesses of Eq.~(\ref{loproc}). Using our normalization, the two
independent (non-vanishing) partonic cross 
sections for the direct component are:
\beq
\label{siqqgamma}
{\hat \sigma}_{q{\bar q} \to g\gamma}^{(0)}(x) = \pi \, 
e_q^2 \,\frac{C_F}{N_c} \;\frac{x^2}{\sqrt {1-x^2}}
\left(2 - x^2\right) 
\eeq
\beq 
\label{siqggamma}
{\hat \sigma}_{qg \to q\gamma}^{(0)}(x) = 
{\hat \sigma}_{{\bar q}g \to {\bar q}\gamma}^{(0)}(x) = \pi \,
e_q^2 \,\frac{1}{2N_c} \;\frac{x^2}{\sqrt {1-x^2}}
\left(1 + \frac{x^2}{4}\right) \;\;,
\eeq
where $e_q$ is the quark electric charge. Note that, having integrated
over the photon pseudorapidity, the expressions 
(\ref{siqqgamma}, \ref{siqggamma}) 
%for ${\hat \sigma}_{ab\to d \gamma}^{(0)}$ 
are even functions of the 
photon transverse energy $E_T$, i.e. they
depend on $x^2$ rather than on $x$. The NLO terms 
${\hat \sigma}_{ab\to \gamma}^{(1)}$ in Eq.~(\ref{pxsg}) were first computed in
Ref.~\cite{aurenche}.

The partonic contributions ${\hat \sigma}_{ab\to c}$
to the fragmentation component of the cross section are exactly equal to those
of the single-hadron inclusive distribution. Their explicit calculation
up to NLO was performed in Ref.~\cite{ACGG}.

We are mainly interested in the behaviour of QCD corrections near the
partonic-threshold region $x \to 1$, i.e. when the transverse energy $E_T$ of 
the photon approaches the partonic centre-of-mass energy $\sqrt {x_1x_2S}$.
In this region, the LO
cross sections (\ref{siqqgamma}, \ref{siqggamma}) behave as
\beq
\label{lobeh}
{\hat \sigma}_{ab\to d \gamma}^{(0)}(x) \sim \frac{1}{\sqrt {1-x^2}} \;\;.
\eeq
This integrable singularity is a typical phase-space effect. 
At higher
perturbative orders, the singularity in Eq.~(\ref{lobeh}) is enhanced
by double-logarithmic corrections due to soft-gluon radiation and the cross
section contributions in Eqs.~(\ref{pxsg}, \ref{pxsc}) behave as
\beq
\label{hobeh}
{\hat \sigma}^{(n)}(x) \sim 
{\hat \sigma}^{(0)}(x) \; \ln^{2n} (1-x) \;\;.
\eeq
Resummation of these soft-gluon effects to all orders in perturbation theory
can be important to improve the reliability of the QCD predictions.

\subsection{$N$-moment space}
\label{Nsec}

The resummation program of soft-gluon contributions has to be carried 
out~\cite{Sterman, CT} in the Mellin-transform space, or $N$-space. 
Working in $N$-space, one can disentangle the soft-gluon effects in the parton
densities from those in the partonic cross section and one can
straightforwardly implement and factorize the kinematic constraints of
energy and longitudinal-momentum conservation. 

It is convenient to consider the Mellin transform $\sigma_{\gamma, \,N}(E_T)$
of the dimensionless hadronic distribution 
$E_T^3 d\sigma_\gamma(x_T,E_T)/d E_T$. The $N$-moments
with respect to $x_T^2$ and at 
fixed $E_T$ are thus defined as follows:
\beq
\label{shn}
\sigma_{\gamma, \,N}(E_T) \equiv \int_0^1 dx_T^2 \;(x_T^2)^{N-1} 
\;E_T^3 \frac{d\sigma_\gamma(x_T,E_T)}{d E_T} \;\;.
\eeq 
In $N$-moment space, Eq.~(\ref{1pxsgamma}) takes a simple factorized form
\beeq
\label{1pxsngamma}
\sigma_{\gamma, \,N}(E_T) &=& \sum_{a,b}
f_{a/H_1,\,N+1}(\mu_F^2) \;f_{b/H_2,\,N+1}(\mu_F^2)
\nonumber \\
&\cdot& \left\{
{\hat \sigma}_{ab\to \gamma, \;N}(\as(\mu^2); E_T^2, \mu^2, \mu_F^2, \mu_f^2) 
\right. \\
&+& \left. \sum_c \,
{\hat \sigma}_{ab\to c, \;N}(\as(\mu^2); E_T^2, \mu^2, \mu_F^2, \mu_f^2)
\;d_{c/\gamma, \,2N+3}(\mu_f^2) \right\} \;\;, \nonumber
\eeeq
where we have introduced the customary $N$-moments $f_{a/H,\,N}$ and 
$d_{a/\gamma, \,N}$ of the parton
densities and parton fragmentation functions:
\beeq 
\label{pdm}
f_{a/H,\,N}(\mu^2) &\equiv& \int_0^1 dx \;x^{N-1} \;f_{a/H}(x,\mu^2) \;\;,
\\
\label{pffm}
d_{a/\gamma, \,N}(\mu^2) &\equiv& \int_0^1 dz \;z^{N-1} 
\;d_{a/\gamma}(z,\mu^2) \;\;.
\eeeq

Note that the $N$-moments of the partonic cross sections in 
Eq.~(\ref{1pxsngamma}) are again defined with respect to $x_T^2$:
\beq
\label{Ndef}
{\hat \sigma}_{ab\to \gamma, \;N}(\as(\mu^2); E_{T}^2, \mu^2, \mu_F^2, \mu_f^2)
\equiv 
\int_0^1 dx^2 \;(x^2)^{N-1} \;
{\hat \sigma}_{ab\to \gamma}(x,\as(\mu^2); E_{T}^2, \mu^2, \mu_F^2, \mu_f^2)
\;\;.
\eeq
In particular, the $N$-moments of the LO contributions in 
Eqs.~(\ref{siqqgamma}, \ref{siqggamma}) are given by the following explicit
expressions:
\beq
\label{qgammaN}
{\hat \sigma}_{q{\bar q}\to g\gamma, \;N}^{(0)} = \pi \,e_q^2 
\,\frac{C_F}{N_c} \;\frac{\Gamma(1/2) \;\Gamma(N+1)}{\Gamma(N+5/2)}
\;(2+N) \;\;,
\eeq
\beq
\label{ggammaN}
{\hat \sigma}_{qg\to q\gamma, \;N}^{(0)} =
{\hat \sigma}_{{\bar q}g\to {\bar q}\gamma, \;N}^{(0)} = \pi \,e_q^2 
\,\frac{1}{8N_c} \;\frac{\Gamma(1/2) \;\Gamma(N+1)}{\Gamma(N+5/2)}
\;(7+5N) \;\;.
\eeq

Note also the pattern of moment indices in the various factors of 
Eq.~(\ref{1pxsngamma}), i.e. $f_{a/H,\,N+1}$ for the parton densities and
$d_{c/\gamma, \,2N+3}$ for the parton fragmentation functions. This
non-trivial pattern follows from the conservation of the longitudinal
and transverse momenta.

The threshold region $x_T \to 1$ corresponds to the limit $N \to \infty$
in $N$-moment space. In this limit, the soft-gluon corrections  
(\ref{hobeh}) to the higher-order contributions of the partonic cross sections
become
\beq
\label{hobehn}
{\hat \sigma}^{(n)}_N \sim 
{\hat \sigma}^{(0)}_N \; \ln^{2n} N \;\;.
\eeq
The resummation of the soft-gluon logarithmic corrections to all orders in
perturbation theory is considered in the following Section.  

\section{Soft-gluon resummation at high $E_T$}
\label{secresum}

\subsection{Resummed cross section to NLL accuracy}
\label{secnll}

In the threshold or large-$N$ limit, the various partonic channels 
contribute in different ways to the prompt-photon cross section
$\sigma_{\gamma, \,N}(E_T)$ in Eq.~(\ref{1pxsngamma}).

Firstly, we can compare the direct and fragmentation contributions to 
Eq.~(\ref{1pxsngamma}).
%Note first that the direct and fragmentation contributions
%${\hat \sigma}_{ab\to \gamma, \;N}$ and ${\hat \sigma}_{ab\to c, \;N}$ to 
The partonic cross sections
${\hat \sigma}_{ab\to \gamma, \;N}$ and ${\hat \sigma}_{ab\to c, \;N}$ 
have the same large-$N$ behaviour, but,
owing to the hard
(although collinear) emission always involved in any splitting process
$c \to \gamma + X$, the photon-fragmentation function $d_{c/\gamma, \,N}$
is of the order of $1/N$. Therefore, in Eq.~(\ref{1pxsngamma})
the fragmentation component is formally suppressed by
a factor of $1/N$ with respect to the direct component and in our resummed
calculation we can neglect the fragmentation contributions.
%Accordingly, the 
%resummed partonic cross sections for the direct processes turn out to be
%independent of the fragmentation scale $\mu_f$.

Then, we can discuss the differences in the large-$N$ behaviour of the partonic
cross sections
${\hat \sigma}_{ab\to \gamma, \;N}(\as)$ for the direct processes.
%; E_T^2, \mu^2, \mu_F^2, \mu_f^2)$. 
The cross sections for the partonic channels
$ab = q{\bar q'}, {\bar q}q', qq, qq', {\bar q}{\bar q}, {\bar q}{\bar q'}$
($q$ and $q'$ denote quarks of different flavours) vanish at LO and are hence
suppressed by a factor of $\as$ with respect to 
${\hat \sigma}_{q{\bar q}\to \gamma, \;N}(\as),
{\hat \sigma}_{qg\to \gamma, \;N}(\as),
{\hat \sigma}_{{\bar q}g\to \gamma, \;N}(\as)$. Moreover, in the 
large-$N$ limit this relative suppression is furtherly enhanced by a factor of
${\cal O}(1/N)$ because the photon has to be accompanied by (at least)
two final-state fermions that are not produced by the decay of an
off-shell gluon. Therefore, we make no attempt to resum soft-gluon
corrections to these partonic channels. The partonic cross section
${\hat \sigma}_{gg\to \gamma, \;N}(\as)$ has a different large-$N$ behaviour.
It begins to contribute at NLO via
the partonic process $g+g \to \gamma + q + {\bar
q}$, which again leads to a suppression effect of ${\cal O}(1/N)$
with respect to the LO subprocesses. However, owing to the photon-gluon
coupling through a fermion box, the partonic subprocess $g+g \to \gamma + g$
is also permitted. This subprocess 
is logarithmically-enhanced by multiple soft-gluon radiation in the 
final state, but it starts to contribute only at 
%the next-to-next-to-leading order 
NNLO in perturbation theory. It follows that the 
partonic channel
$ab = gg$ is suppressed by a factor of $\as^2$ with respect to the LO
partonic channels $ab = q{\bar q}, qg, {\bar q}g$ and it enters the resummed
cross section only at NNLL accuracy.

In conclusion, since we are interested in explicitly perform
soft-gluon resummation up to NLL order, we can limit ourselves to
considering the partonic cross sections
${\hat \sigma}_{q{\bar q}\to \gamma}, {\hat \sigma}_{qg\to \gamma},
{\hat \sigma}_{{\bar q}g\to \gamma}$.

As discussed in Sect.~\ref{secgen}, the soft-gluon
%logarithmically-enhanced threshold
corrections to the partonic channels $ab = q{\bar q}, qg, {\bar q}g$
are not affected by colour correlations. Thus, in the resummed
expressions ${\hat \sigma}_{ab\to \gamma, \;N}^{({\rm res})}$ for the
partonic cross sections, the logarithmically-enhanced threshold contributions
can be factorized with respect to the corresponding LO cross sections
${\hat \sigma}_{ab\to d\gamma, \;N}^{(0)}$ in 
Eqs.~(\ref{qgammaN}, \ref{ggammaN}).
The {\em all-order} resummation formulae are
\beeq
%\label{gammaresqq}
{\hat \sigma}_{q{\bar q}\to \gamma, \;N}^{({\rm res})}(\as(\mu^2); 
E_T^2, \mu^2, \mu_F^2, \mu_f^2) &=& \alpha \;\as(\mu^2) 
\;{\hat \sigma}_{q{\bar q}\to g\gamma, \;N}^{(0)} 
\;C_{q{\bar q} \to \gamma}(\as(\mu^2),Q^2/\mu^2;Q^2/\mu_F^2) \nonumber \\
\label{gammaresqq}
&\cdot&
\Delta_{N+1}^{q{\bar q} \to g \gamma}(\as(\mu^2),Q^2/\mu^2;Q^2/\mu_F^2)
\;\;, \\
%\eeeq 
%\beeq
%\label{gammaresqg}
{\hat \sigma}_{qg\to \gamma, \;N}^{({\rm res})}(\as(\mu^2); 
E_T^2, \mu^2, \mu_F^2, \mu_f^2) &=& \alpha \;\as(\mu^2) 
\;{\hat \sigma}_{qg\to q\gamma, \;N}^{(0)} 
\;C_{qg \to \gamma}(\as(\mu^2),Q^2/\mu^2;Q^2/\mu_F^2) \nonumber \\
\label{gammaresqg}
&\cdot& 
\Delta_{N+1}^{qg \to q \gamma}(\as(\mu^2),Q^2/\mu^2;Q^2/\mu_F^2)
\;\;, \\
%\eeeq
%\beeq
\label{gammaresqbarg}
{\hat \sigma}_{{\bar q}g\to \gamma, \;N}^{({\rm res})}(\as(\mu^2); 
E_T^2, \mu^2, \mu_F^2, \mu_f^2) &=& 
{\hat \sigma}_{qg\to \gamma, \;N}^{({\rm res})}(\as(\mu^2); 
E_T^2, \mu^2, \mu_F^2, \mu_f^2)
\;\;,
\eeeq 
where 
%$Q^2 = 2 E_T^2$.
\beq
\label{etscale}
Q^2 = 2 E_T^2 \;\;.
\eeq

Note that the right-hand side of Eqs.~(\ref{gammaresqq}, \ref{gammaresqg})
does not depend on the factorization scale $\mu_f$ of the photon fragmentation
functions. Thus, the resummed partonic cross sections
${\hat \sigma}_{ab\to \gamma, \;N}^{({\rm res})}$ turn out to be
independent of $\mu_f$. This is in agreement with the subdominance
of the fragmentation contributions near threshold, as discussed above.

The functions $C_{ab\to \gamma}(\as)$ in 
Eqs.~(\ref{gammaresqq}, \ref{gammaresqg}) do not depend on $N$ and, thus, 
contain all the contributions that are constant in the large-$N$ limit.
These functions are computable as power series expansions in $\as$
\beq
\label{cgamma}
C_{ab\to \gamma}(\as(\mu^2),Q^2/\mu^2;Q^2/\mu_F^2) = 
1 + \sum_{n=1}^{+\infty} \; 
\left( \frac{\as(\mu^2)}{\pi} \right)^n \;
C_{ab\to \gamma}^{(n)}(Q^2/\mu^2;Q^2/\mu_F^2) \;\;.
\eeq
The physical origin and the structure of the constant factors 
$C_{ab\to \gamma}(\as)$ is discussed in Sect.~\ref{secconfac}.

The $\ln N$-dependence of the resummed cross sections 
is entirely embodied by the radiative factors
$\Delta_N^{ab \to d \gamma}$ on the right-hand side of 
Eqs.~(\ref{gammaresqq}, \ref{gammaresqg}). 
Note, the mismatch between the moment index of the radiative factor and that
of ${\hat \sigma}_{ab\to d\gamma, \;N}^{(0)}$: the former depends on $N+1$, 
like the parton densities in Eq.~(\ref{1pxsngamma}). The explicit expressions 
of the radiative factors are given in the following subsection.

\subsection{The radiative factors}
\label{secradfac}

The soft-gluon factors $\Delta_N^{ab \to d \gamma}$ depend on the flavour
of the QCD partons $a, b, d$ involved in the LO hard-scattering
subprocess $a+b \to d + \gamma$. According to the discussion of 
Sect.~\ref{secgen}, the resummed expressions
for $\Delta_N^{ab \to d \gamma}$ have an exponential form.
To explain the exponentiation structure 
and to facilitate the comparison with other hadroproduction processes,
we use a notation similar to that in Ref.~\cite{CT} and we write the
prompt-photon radiative factors as follows
\beeq
\label{deltangamma}
\Delta_N^{ab \to d \gamma}(\as(\mu^2),Q^2/\mu^2;Q^2/\mu_F^2) &=&
\Delta_N^a(\as(\mu^2),Q^2/\mu^2;Q^2/\mu_F^2)
\; \Delta_N^b(\as(\mu^2),Q^2/\mu^2;Q^2/\mu_F^2) 
\nonumber \\
&\cdot& J_N^d(\as(\mu^2),Q^2/\mu^2) \;
\Delta_N^{({\rm int}) \,ab \to d \gamma}(\as(\mu^2),Q^2/\mu^2) \;\;.
\eeeq
The resummed formulae to NLL accuracy for the various contributions
on the right-hand side of this equation are presented below.
 
Each term $\Delta_N^a(\as(\mu^2),Q^2/\mu^2;Q^2/\mu_F^2)$ depends on the 
flavour $a$ of a single parton,
on the factorization scheme of the parton density $f_{a/H, \,N}(\mu_F^2)$ and 
on the factorization scale $\mu_F$. In the $\MSB$ scheme, 
we have
\beq
\label{deltams}
\Delta_N^a(\as(\mu^2),Q^2/\mu^2;Q^2/\mu_F^2) = \exp \left\{
\int_0^1 dz \;\frac{z^{N-1} -1}{1-z} \;
\int_{\mu_F^2}^{(1-z)^2Q^2} 
\frac{dq^2}{q^2} \;A_a(\as(q^2)) 
%\nonumber \\ &+& 
+ {\cal O}(\as(\as \ln N)^k) \right\} \,,
\eeq
where $A_a(\as)$ are perturbative functions
\beq
\label{Afun}
A_a(\as)={\asp} A_a^{(1)}+\left(\asp \right)^2 A_a^{(2)} + {\cal O}(\as^3)
\;\;.
\eeq 
The lower-order terms $A_a^{(1)}$ and $A_a^{(2)}$ are 
\beq
\label{A12coef}
A_a^{(1)}= C_a\;,\;\;\;\; A_a^{(2)}=\frac{1}{2} \; C_a K \;,
\eeq
where $C_a = C_F$ if $a=q,{\bar q}$ and $C_a = C_A$ if $a=g$, while
the coefficient $K$ is the same both for quarks \cite{KT} and for gluons 
\cite{CdET}
and it is given by\footnote{In $SU(N_c)$ QCD, the colour factors are
$C_F= (N_c^2-1)/(2N_c), C_A= N_c$ and $T_R=1/2$.}
\beq
\label{kcoef}
K = C_A \left( \frac{67}{18} - \frac{\pi^2}{6} \right) 
- \frac{10}{9} T_R N_f \;.
\eeq

The term $J_N^d(\as(\mu^2),Q^2/\mu^2)$
depends on the parton 
flavour $d$ and is independent both of the factorization scale and of the 
factorization scheme:
\beeq
J_N^d(\as(\mu^2),Q^2/\mu^2) &=& \exp \left\{
\int_0^1 dz \;\frac{z^{N-1} -1}{1-z} 
\;\left[ \int_{(1-z)^2Q^2}^{(1-z)Q^2} 
\frac{dq^2}{q^2} \;A_d(\as(q^2)) +  \frac{1}{2} \;B_d(\as((1-z)Q^2)) \right]
\right. \nonumber \\ 
\label{jfun}
&+& \left. \frac{}{} {\cal O}(\as(\as \ln N)^k) \right\} \;\;.
\eeeq
The functions $A_d(\as)$ are given in Eq.~(\ref{Afun}) and
the functions $B_d(\as)$ have analogous perturbative expansions:
\beq
\label{Bfun}
B_d(\as)={\asp} B_d^{(1)} + {\cal O}(\as^2)
\eeq
with \cite{KT, CdET}
\beq
\label{B1coef}
B_{d=q,{\bar q}}^{(1)}= - \frac{3}{2} C_F \;,\;\;\;\;\;
B_{d=g}^{(1)}= - \frac{1}{6} \;( 11 C_A - 4 T_R N_f ) \;\;.
\eeq

Likewise $J_N^d$, the remaining contribution 
$\Delta_N^{({\rm int})}$ in Eq.~(\ref{deltangamma}) is independent of
the factorization scale and scheme. Nonetheless, it depends on the flavours of
all the QCD partons entering the LO scattering subprocess:
\beq
\label{deltaintgamma}
\Delta_N^{({\rm int}) \,ab \to d \gamma}(\as(\mu^2),Q^2/\mu^2) = \exp \left\{
\int_0^1 dz  \;\frac{z^{N-1} -1}{1-z} \; 
\;D_{ab \to d \gamma}(\as((1-z)^2Q^2)) 
+ {\cal O}(\as(\as \ln N)^k) \right\}\;\;.
\eeq
The function $D_{ab \to d \gamma}(\as)$ has the following perturbative
expansion
\beq
\label{Dgammafun}
D_{ab \to d \gamma}(\as) = {\asp} D_{ab \to d \gamma}^{(1)} + {\cal O}(\as^2)
\;\;,
\eeq
with
\beq
\label{D1coef}
D_{ab \to d \gamma}^{(1)} = \left( C_a + C_b - C_d \right) \ln 2 \;\;.
\eeq

The factorized structure in Eq.~(\ref{deltangamma}) has a direct physical
interpretation. The factors $\Delta_N^a$ and $\Delta_N^b$ take into account
soft-gluon radiation emitted collinearly to the {\em initial}-state
partons. Consistently, these are the sole factors that depend on the
factorization scale $\mu_F$ of the parton densities of the colliding hadrons.
The factor $J_N^d$ is due to collinear (either soft or hard) radiation
in the {\em final}-state jet that is produced by the fragmentation of the 
parton $d$ recoiling against the triggered photon. The factor 
$\Delta_N^{({\rm int})}$ contains the contribution of soft-gluon emission at 
large angle with respect to the direction of the hard partons entering the
LO scattering subprocess. This factor thus embodies the soft-gluon
{\em interference} effects anticipated in Sect.~\ref{secgen}.

According to this interpretation, the perturbative functions in 
Eqs.~(\ref{Afun}, \ref{Bfun}, \ref{Dgammafun}) measure the intensity
of the coupling of $i)$~{\em soft-collinear} gluons (function $A_a(\as)$),
$ii)$~{\em hard-collinear} partons (function $B_a(\as)$) and
$iii)$~{\em large-angle soft} gluons (function $D_{ab \to d \gamma}(\as)$).
Note that, due to their collinear nature, the functions $A_a(\as)$ and
$B_a(\as)$ depend on the colour and flavour of the sole parton a.
On the contrary, $D_{ab \to d \gamma}(\as)$ depends on the colour charges
of all the QCD partons.

The physical origin of the several contributions on the right-hand side of
Eq.~(\ref{deltangamma}) is furtherly discussed in Sect.~\ref{seccomp}, where
we compare the promp-photon radiative factors with the analogous resummed
factors that control the threshold behaviour of other hadroproduction
processes. In the rest of this section we limit ourselves to comment
on few additional features of the resummed contributions to the prompt-photon
cross section.

The various factors in Eq.~(\ref{deltangamma}) contribute to the resummed
prompt-photon cross section at different level of logarithmic accuracy.
If we simply consider the double-logarithmic (DL) approximation, which consists
in resumming only the terms $\as^n \ln^{2n}N$, we can neglect the interference
factor $\Delta_N^{({\rm int})}$ and the $B(\as)$ function in Eq.~(\ref{jfun})
and we can expand the exponent in $\Delta_N^a$ and $J_N^d$ to its
first order in $\as$:
\beeq
\label{dldeltams}
\Delta_N^a(\as(\mu^2),Q^2/\mu^2;Q^2/\mu_F^2) &\simeq&
\exp \left\{ + \, 2C_a \frac{\as}{2\pi} \ln^2 N \right\} \;\;, \\
\label{dljfun}
J_N^d(\as(\mu^2),Q^2/\mu^2) &\simeq&
\exp \left\{ - \, C_d \frac{\as}{2\pi} \ln^2 N \right\} \;\;.
\eeeq
The complete set of LL terms is obtained by neglecting the functions
$B(\as),D(\as)$ in Eqs.~(\ref{jfun}, \ref{deltaintgamma}), by truncating
$A_a(\as)$ to their first order and using the LO running of the coupling
$\as(q^2)$. At the NLL order, also the contribution of the coefficients
$A_a^{(2)}, B_a^{(1)}$ and $D_{ab \to d \gamma}^{(1)}$ has to be included.

Note that different scales, e.g. $q^2, (1-z)^2Q^2, (1-z)Q^2$, appear on
the right-hand sides of Eqs.~(\ref{deltams}, \ref{jfun}, 
\ref{deltaintgamma}). In particular, the scales in the 
$q^2$-integration limits of Eq.~(\ref{deltams}) are different from those
of Eq.~(\ref{jfun}), and the $B$ function in Eq.~(\ref{jfun}) depends
on $\as((1-z)Q^2)$ while the $D$ function in the interference contribution
(\ref{deltaintgamma}) depends on $\as((1-z)^2Q^2)$. These scales follows from
the hard-scattering kinematics, which affect in a different way initial- or
final-state emission and collinear or soft radiation. 

Note, also, that the
renormalization scale $\mu$ does not explicitly enter the right-hand side
of Eqs.~(\ref{deltams}, \ref{jfun}, \ref{deltaintgamma}). This is because the
radiative factors are renormalization-group-invariant quantities when evaluated
to all order in perturbation theory. Only when the all-order expressions are 
truncated to a certain degree of logarithmic accuracy, the
renormalization-scale dependence explicitly appears as a higher-order effect.

Since we know the radiative factors only to NLL order, we use the 
Eqs.~(\ref{deltams}, \ref{jfun}, \ref{deltaintgamma}) by replacing
$\as(k^2)$ (with $k^2=q^2, (1-z)^2Q^2, (1-z)Q^2)$ with its NLO expansion
in terms of $\as(\mu^2)$ and $k^2$ (cf. Appendix~A), and we explicitly carry
out the $z$ and $q^2$ integrals by neglecting terms beyond NLL accuracy. We
thus write the prompt-photons radiative factors as follows:
\beeq
\Delta_N^{ab \to d \gamma}\!\left(\as(\mu^2),\frac{Q^2}{\mu^2};
\frac{Q^2}{\mu_F^2}\right) &=&
\exp \left\{ \ln N \; g_{ab}^{(1)}(b_0\as(\mu^2)\ln N) +
g_{ab}^{(2)}(b_0\as(\mu^2)\ln N,Q^2/\mu^2;Q^2/\mu_F^2 ) \right. \nonumber \\
\label{deltanll}
&+& \left. {\cal O}(\as(\as \ln N)^k) \right\} \;\;.
\eeeq
The functions $g^{(1)}$ and $g^{(2)}$ resum the LL and NLL terms, respectively.
These functions are different for the $q{\bar q}$ and $qg$ partonic channels
of Eqs.~(\ref{gammaresqq}) and (\ref{gammaresqg}), and are 
explicitly computed in Appendix~A. We find:
\beq
\label{g1fun}
g_{q{\bar q}}^{(1)}(\lambda) = (2C_F - C_A) \;h^{(1)}(\lambda) +
C_A \;h^{(1)}(\lambda/2) \;, \;\;\; \;\;
g_{qg}^{(1)}(\lambda) = C_A \;h^{(1)}(\lambda) +
C_F \;h^{(1)}(\lambda/2) \;, 
\eeq
and
\beeq
\label{g2qq}
g_{q{\bar q}}^{(2)}\!\left(\lambda,\frac{Q^2}{\mu^2};\frac{Q^2}{\mu_F^2}\right)
&=& (2C_F - C_A) \;h^{(2)}(\lambda) + 2 \,C_A \;h^{(2)}(\lambda/2)  \\
&+& \frac{2C_F - C_A}{2\pi b_0} \ln 2 \ln(1-2\lambda) 
+ \frac{ C_A \GE -  \pi b_0}{\pi b_0} \ln(1-\lambda) 
- \frac{2C_F}{\pi b_0} \;\lambda \ln \frac{Q^2}{\mu_F^2}
\nonumber \\
&+& \left\{ \frac{C_F}{\pi b_0} \Bigl[ 2\lambda + \ln(1-2\lambda) \Bigr]  
+ \frac{C_A}{2\pi b_0} \Bigl[ 2 \ln(1-\lambda) - \ln(1-2\lambda) \Bigr]
\right\} \ln \frac{Q^2}{\mu^2} \;, \nonumber
\eeeq
\beeq
\label{g2qg}
g_{qg}^{(2)}\!\left(\lambda,\frac{Q^2}{\mu^2};\frac{Q^2}{\mu_F^2}\right)
&=& C_A \;h^{(2)}(\lambda) + 2 \,C_F \;h^{(2)}(\lambda/2) \\
&+& \frac{C_A}{2\pi b_0} \ln 2 \ln(1-2\lambda) 
+ \frac{4 C_F \,\GE - 3 C_F}{4\pi b_0} \ln(1-\lambda) 
- \frac{C_F + C_A}{\pi b_0} \;\lambda \ln \frac{Q^2}{\mu_F^2}
\nonumber \\
&+& \left\{ \frac{C_F + C_A}{2\pi b_0} \Bigl[ 2\lambda + \ln(1-2\lambda) \Bigr]  
+ \frac{C_F}{2\pi b_0} \Bigl[ 2 \ln(1-\lambda) - \ln(1-2\lambda) \Bigr]
\right\} \ln \frac{Q^2}{\mu^2} \;, \nonumber
\eeeq
where $\GE = 0.5772\ldots$ is the Euler number and 
%where 
$b_0, b_1$ are the first two coefficients of the QCD $\beta$-function
\beq
\label{betass}
b_0 = \frac{11 C_A - 4 T_R N_f}{12\pi}\;,\;\;\;\;\; b_1 =
\frac{17 C_A^2 - 10 C_A T_R N_f -6 C_F T_R N_f}{24\pi^2}\;.
\eeq
The auxiliary functions $h^{(1)}$ and $h^{(2)}$ in Eqs.~(\ref{g1fun}) and
(\ref{g2qq}, \ref{g2qg}) are 
\beq
\label{htll}
h^{(1)}(\lambda) =
\frac{1}{2\pi b_0 \lambda}\Bigl[ 2\lambda + (1-2\lambda)
\ln(1-2\lambda) \Bigr] \; , 
\eeq
\beq
\label{htnll}
h^{(2)}(\lambda) =
\frac{b_1}{2\pi b_0^3}\Bigl[ 2\lambda + \ln(1-2\lambda)
+\half\ln^2(1-2\lambda) \Bigr] 
- \frac{\GE}{\pi b_0} \ln(1-2\lambda) 
- \frac{K}{4\pi^2 b_0^2}\Bigl[ 2\lambda + \ln(1-2\lambda)\Bigr] 
\; , 
\eeq
where $K$ is the coefficient in Eq.~(\ref{kcoef}).

The results in Eqs.~(\ref{deltanll}--\ref{g2qg})
provide us with a theoretical description of soft-gluon resummation in
prompt-photon hadroproduction
at the same level of accuracy as for other hadroproduction processes,
such as the production of Drell-Yan pairs \cite{Sterman, CT} or heavy quarks
\cite{kidon,BCMN}. These results can be used for detailed quantitative studies
along the lines of Refs.~\cite{CMNT2, BCMN}. In this paper we do not present
numerical analyses and we limit ourselves to discuss the expected sign and size
of the resummation effects.

In the near-threshold region, radiation in the final state is kinematically
inhibited. On physical basis, one thus expects that resummation of the
ensuing logarithmically-enhanced corrections produces suppression of the
cross section. This argument applies to {\em hadronic} cross sections, but
it is not necessarely valid for {\em partonic} cross sections. The partonic
cross section is what is left after factorization of long-distance physics
into the parton distributions. Since all-order resummation is in part
automatically implemented in the definition of the parton densities, 
the remaining resummation effects can either enhance or deplete
the partonic cross section. 

Among the various terms on the right-hand side of Eq.~(\ref{deltangamma}),
some factors are smaller and some others are larger than unity.
The exponent of the initial-state contribution $\Delta_N^a$ in 
Eq.(\ref{deltams}) is positive definite and, hence, $\Delta_N^a >> 1$ when
$N \to \infty$. The presence of this `anti-Sudakov' form factor is typical
of partonic cross sections that are evaluated by factorizing parton
densities defined in the $\MSB$ factorization scheme. In the case of the
final-state contribution $J_N^d$, no additional factorization has been
performed. Therefore, when $N \to \infty$ the exponent in Eq.~(\ref{jfun})
is negative and $J_N^d \ll 1$ is a `true' Sudakov form factor, as naively
expected. The sign of the exponent in Eq.~(\ref{deltaintgamma}) is not definite
($D_{q{\bar q} \to g \gamma}^{(1)} < 0, 
D_{q g \to q \gamma}^{(1)}= D_{{\bar q} g \to {\bar q} \gamma}^{(1)} > 0$)
as expected for an interference term. However, the contribution of 
$\Delta_N^{({\rm int}) \,ab \to d \gamma}$ is subleading with respect to those
of $\Delta_N^a$ and $J_N^d$.

From the overall inspection of the effect of the radiative-factor
contributions to Eq.~(\ref{deltangamma}), we infer that,  
in the case of prompt-photon production, the resummed
partonic cross sections 
${\hat \sigma}_{q{\bar q}\to \gamma, \;N}^{({\rm res})}$
and ${\hat \sigma}_{qg\to \gamma, \;N}^{({\rm res})}$
in Eqs.~(\ref{gammaresqq}) and (\ref{gammaresqg}) are both enhanced
with respect to their LO approximations 
${\hat \sigma}_{q{\bar q}\to g\gamma, \;N}^{(0)}$,
${\hat \sigma}_{qg\to q\gamma, \;N}^{(0)}$. Moreover, the enhancement in the
$qg$ partonic channel is larger than that in the $q{\bar q}$ channel.

This conclusion can also be argued by a
simplified treatment within the DL approximation.
Inserting Eqs.~(\ref{dldeltams}, \ref{dljfun}) into Eq.~(\ref{deltangamma}),
we obtain
\beeq
\label{dlresqg'}
{\hat \sigma}_{qg\to \gamma, \;N}^{({\rm res})} &\simeq&
{\hat \sigma}_{qg\to q\gamma, \;N}^{(0)} \;
\exp \left\{ \Bigl[ 2C_F + 2C_A - C_F \Bigr]
\frac{\as}{2\pi} \ln^2 N \right\}  \\
\label{dlresqg}
&=& {\hat \sigma}_{qg\to q\gamma, \;N}^{(0)} \;
\exp \left\{ \left( C_F + 2C_A \right)
\frac{\as}{2\pi} \ln^2 N \right\} > {\hat \sigma}_{qg\to q\gamma, \;N}^{(0)} 
\;\;, \\
&~& \nonumber \\
\label{dlresqq'}
{\hat \sigma}_{q{\bar q}\to \gamma, \;N}^{({\rm res})} &\simeq&
{\hat \sigma}_{q{\bar q}\to q\gamma, \;N}^{(0)} \;
\exp \left\{ \left[ 2C_F + 2C_F - C_A \right]
\frac{\as}{2\pi} \ln^2 N \right\}  \\
\label{dlresqq}
&=& {\hat \sigma}_{q{\bar q}\to q\gamma, \;N}^{(0)} \;
\exp \left\{ \left( 4C_F - C_A \right)
\frac{\as}{2\pi} \ln^2 N \right\} > 
{\hat \sigma}_{q{\bar q}\to q\gamma, \;N}^{(0)} 
\;\;, \\
&~& \nonumber \\
\label{dlrat}
\frac{{\hat \sigma}_{qg\to \gamma, \;N}^{({\rm res})}}
{{\hat \sigma}_{q{\bar q}\to \gamma, \;N}^{({\rm res})}} &\simeq&
\frac{{\hat \sigma}_{qg\to q\gamma, \;N}^{(0)}}
{{\hat \sigma}_{q{\bar q}\to q\gamma, \;N}^{(0)}} \;
\exp \left\{ 3(C_A - C_F) 
\frac{\as}{2\pi} \ln^2 N \right\} >
\frac{{\hat \sigma}_{qg\to q\gamma, \;N}^{(0)}}
{{\hat \sigma}_{q{\bar q}\to q\gamma, \;N}^{(0)}} \;\;.
\eeeq
The first, second and third terms in the square bracket on the right-hand side
of Eqs.~(\ref{dlresqg'}, \ref{dlresqq'}) are respectively due to the 
initial-state factors $\Delta_N^a$, $\Delta_N^b$ and to the final-state factor
$J_N^d$. Note that, for a definite parton $a$, the initial-state enhancement
$\Delta_N^a$ is larger than the final-state suppression $J_N^a$
(see the difference by a factor of two in the exponent of
Eqs.~(\ref{dldeltams}, \ref{dljfun})). In the $qg$ channel the 
final-state contribution $\ln J_N^q$ is thus overcompensated by 
$\ln \Delta_N^q$ and
this leads to the enhancement in Eq.~(\ref{dlresqg}). In the $q{\bar q}$
channel, instead, it is the total initial-state contribution
$(\ln \Delta_N^q + \ln \Delta_N^{\bar q})$ 
that, owing to the colour-charge relation $C_F \sim C_A/2$,
overcompensates $\ln J_N^g$. Finally, the enhancement in Eq.~(\ref{dlrat})
is simply due the fact that the gluon colour charge $C_A$ is larger that the 
quark charge $C_F$ and, thus, $\Delta_N^g > \Delta_N^q$ and $J_N^q > J_N^g$.  

Note that this conclusion directly applies only to the asymptotic limit
$N \to \infty$ or $E_T \to {\sqrt S}/2$. In the case of kinematic configurations
of experimental interest, subleading effects and their dependence on the
$x$-shape of the parton distributions and on the renormalization and
factorization scale have to be carefully estimated.

\subsection{The constant factors}
\label{secconfac}

Expanding the resummed expressions in Eqs.(\ref{deltanll}--\ref{g2qg}) to the 
first order in $\as$ and using Eqs.~(\ref{gammaresqq}, \ref{gammaresqg}),
we obtain
%\beeq
%&&{\hat \sigma}_{q{\bar q}\to \gamma, \;N}^{({\rm res})}(\as(\mu^2); 
%E_T^2, \mu^2, \mu_F^2, \mu_f^2) = \; \alpha
%\;{\hat \sigma}_{q{\bar q}\to g\gamma, \;N}^{(0)}
%\;\as(\mu^2) \left\{ 1 + \frac{\as(\mu^2)}{\pi}
%\left[ \left( 2C_F - \frac{1}{2} C_A \right) \ln^2 N \right. \right. 
%\nonumber \\
%&&+ \!\left. \left. \Bigl( \GE (4 C_F - C_A) - (2 C_F - C_A) \ln 2
%+ \pi b_0 - 2 C_F \ln \frac{2E_T^2}{\mu_F^2} \;\Bigr) \ln N \right. \right.
%\nonumber \\
%&& + \left. \left.
%C_{q{\bar q}\to \gamma}^{(1)}(Q^2/\mu^2;Q^2/\mu_F^2) \right]  
%+  \frac{}{} {\cal O}(\as^2) \right\}
%\eeeq
\beeq
{\hat \sigma}_{q{\bar q}\to \gamma, \;N}^{({\rm res})}(\as(\mu^2); 
E_T^2, \mu^2, \mu_F^2, \mu_f^2) &\!\!=\!\!&  \alpha
\;{\hat \sigma}_{q{\bar q}\to g\gamma, \;N}^{(0)}
\;\as(\mu^2) \left\{ 1 + \frac{\as(\mu^2)}{\pi}
\left[ \left( 2C_F - \frac{1}{2} C_A \right) \ln^2 N \right. \right. 
\nonumber \\
&\!\!+\!\!& \left. \left. \Bigl( \GE (4 C_F - C_A) - (2 C_F - C_A) \ln 2
+ \pi b_0 - 2 C_F \ln \frac{2E_T^2}{\mu_F^2} \;\Bigr) \ln N \right. \right.
\nonumber \\
\label{qq1st}
&\!\!+\!\!&  \left. \left.
C_{q{\bar q}\to \gamma}^{(1)}(2E_T^2/\mu^2;2E_T^2/\mu_F^2) \frac{}{} \right]  
+  {\cal O}(\as^2) \right\} \;\;,
\eeeq
\beeq
{\hat \sigma}_{qg\to \gamma, \;N}^{({\rm res})}(\as(\mu^2); 
E_T^2, \mu^2, \mu_F^2, \mu_f^2) &\!\!=\!\!&  \alpha
\;{\hat \sigma}_{qg\to q\gamma, \;N}^{(0)}
\;\as(\mu^2) \left\{ 1 + \frac{\as(\mu^2)}{\pi}
\left[ \left( \frac{1}{2} C_F + C_A \right) \ln^2 N \right. \right. 
\nonumber \\
&\!\!+\!\!& \left. \left. \Bigl( \GE ( C_F + 2 C_A) - C_A \ln 2
+ \frac{3}{4} C_F - ( C_F + C_A ) \ln \frac{2E_T^2}{\mu_F^2} \;\Bigr) 
\ln N \right. \right.
\nonumber \\
\label{qg1st}
&\!\!+\!\!&  \left. \left.
C_{qg\to \gamma}^{(1)}(2E_T^2/\mu^2;2E_T^2/\mu_F^2) \frac{}{} \right]  
+   {\cal O}(\as^2) \right\} \;\;.
\eeeq
One can easily check that the logarithmic terms in
these perturbative expansions agree with those that can be derived
from the complete NLO analytic results of
Refs.~\cite{aurenche, contogouris, gordon93}. From this comparison we can
also extract the first-order constant coefficients 
$C_{q{\bar q}\to \gamma}^{(1)}$ and $C_{qg\to \gamma}^{(1)}$. We find
\beeq
C_{q{\bar q}\to \gamma}^{(1)}(Q^2/\mu^2;Q^2/\mu_F^2) &=& 
\GE^2 \Bigl( 2 C_F - \frac{1}{2} C_A \Bigr)
+ \GE \Bigl[ \pi b_0 - (2 C_F - C_A) \ln 2 \Bigl] 
- \, \frac{1}{2} (2 C_F - C_A) \ln 2 \nonumber \\
\label{cqq1coef}
&+& \frac{1}{2} \,K - K_q 
+ \frac{\pi^2}{3} \Bigl( 2 C_F - \frac{1}{2} C_A \Bigr)
+ \frac{5}{4} (2 C_F - C_A) \ln^2 2 \\
&-& \Bigl( 2 \GE C_F - \frac{3}{2} C_F \Bigr) \ln \frac{Q^2}{\mu_F^2}
- \pi b_0 \ln \frac{Q^2}{\mu^2} \;\;, \nonumber
\eeeq
\beeq
C_{qg\to \gamma}^{(1)}(Q^2/\mu^2;Q^2/\mu_F^2) &=&
\GE^2 \Bigl( \frac{1}{2} C_F + C_A \Bigr)
+ \GE \Bigl[ \frac{3}{4} C_F -  C_A \ln 2 \Bigl]
- \, \frac{1}{10} ( C_F - 2 C_A) \ln 2 \nonumber \\
\label{cqg1coef}
&-& \frac{1}{2} \,K_q 
+ \frac{\pi^2}{60} \Bigl( 2 C_F + 19 C_A \Bigr)
+ \frac{1}{2} C_F  \ln^2 2 \\
&-& \Bigl( \GE (C_F + C_A) - \frac{3}{4} C_F - \pi b_0 \Bigr) 
\ln \frac{Q^2}{\mu_F^2}
- \pi b_0 \ln \frac{Q^2}{\mu^2} \;\;, \nonumber
\eeeq
where 
\beq
\label{kcoeffq}
K_q = \left( \frac{7}{2} - \frac{\pi^2}{6} \right) \,C_F \;\;,
\eeq
and the coefficient $K$ is given in Eq.~(\ref{kcoef}).

The first-order coefficient $C_{ab\to \gamma}^{(1)}$ and, indeed, all the
perturbative coefficients of the constant ($N$-independent) function
$C_{ab\to \gamma}(\as)$ in Eq.~(\ref{cgamma}) are produced by 
hard {\em virtual} contributions and by subdominant (non-logarithmic)
soft corrections to the LO hard-scattering subprocesses.
In both cases the structure of the external hard partons is the same as at LO.
This justifies the all-order factorization of $C_{ab\to \gamma}(\as)$
with respect to ${\hat \sigma}_{ab \to d \gamma, \,N}^{(0)}$ and to
the radiative factor in the resummed partonic
cross sections (\ref{gammaresqq}, \ref{gammaresqg}).

The inclusion of the $N$-independent function $C_{ab\to \gamma}(\as)$
in the resummed formulae does not affect the shape of the cross section
near threshold, but improves the soft-gluon resummation by fixing the
overall (perturbative) normalization of the logarithmic radiative factor.

We can explicitly show \cite{CTTW, BCMN} 
the theoretical improvement that is obtained by combining
the NLL radiative factor with the first-order coefficient $C^{(1)}$.
Expanding the resummation formulae (\ref{gammaresqq}, \ref{gammaresqg}) in
towers of logarithmic contributions as in Eq.~(\ref{genlog}), we have
\beeq
{}\!\!\!\ {}\!\! {\hat \sigma}_N^{({\rm res})}(\as;E_T^2,\mu^2,\mu_F^2) = 
\alpha \, \as \, 
{\hat \sigma}_N^{(0)}&&\!\!\!\! \!\!\!\! \! 
\left\{ 1 + \sum_{n=1}^{\infty} \as^n \left[ c_{n,2n} \;\ln^{2n} N 
+ c_{n,2n-1}(E_T^2/\mu_F^2)  \;\ln^{2n-1} N \right.
\right. \nonumber \\
\label{sigtow}
&&\!+ \left. \left.  c_{n,2n-2}(E_T^2/\mu_F^2,E_T^2/\mu^2) \;\ln^{2n-2} N
+ {\cal O}(\ln^{2n-3} N) \right] \right\} \;,
\eeeq
where $\as= \as(\mu^2)$. The dominant and next-to-dominant
coefficients $c_{n,2n}$ and $c_{n,2n-1}$ are controlled
by evaluating the radiative factor to NLL accuracy. When the NLL radiative
factor is supplemented with the coefficient $C^{(1)}$, we can correctly
control also the coefficients $c_{n,2n-2}$. 
In particular, we can predict (see Appendix~B) the large-$N$ behaviour of
the 
%next-to-next-to-leading order 
NNLO cross sections 
${\hat \sigma}_{ab\to \gamma}^{(2)}$ in Eq.~(\ref{pxsg}) up to 
${\cal O}(\ln N)$. 

Note also that coefficients $c_{n,2n}$ are scale independent and the
coefficients $c_{n,2n-1}$ depend on the sole factorization scale 
$\mu_F$. In the tower expansion (\ref{sigtow}), the first terms that
explicitly depend on the renormalization scale $\mu$ (and on $\mu_F$, as well)
are those controlled by $c_{n,2n-2}$. Their dependence on $\mu$ is
obtained by combining that of $C^{(1)}(E_T^2/\mu_F^2,E_T^2/\mu^2)$
with that of the radiative factor at NLL order. The inclusion of the 
first-order constant coefficient $C^{(1)}$ thus (theoretically) stabilizes
the resummed partonic cross section with respect to variations of
the renormalization scale.
\setcounter{footnote}{0}
\section{Comparison with other processes: soft-gluon interferences and
QCD coherence}
\label{seccomp}
Further insight on the underlying physics mechanism that leads to the
resummed expressions (\ref{gammaresqq}, \ref{gammaresqg}) can be obtained by 
comparing prompt-photon production with other hard-scattering processes.

\begin{figure}[t]
\centerline{\epsfig{file=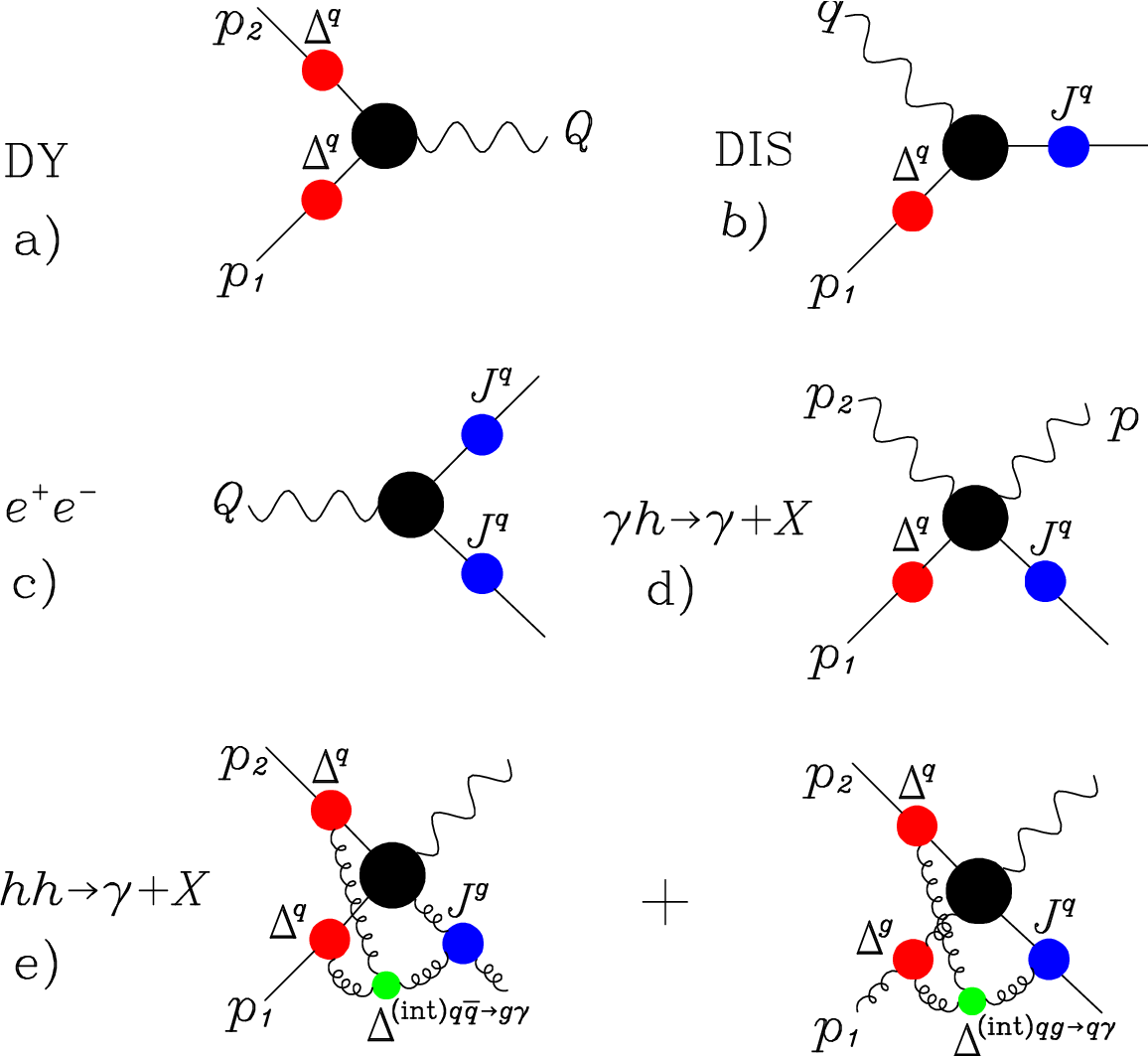,width=12cm,clip=}}
\ccaption{}{\label{fig} Schematic representation of the structure of the
soft-radiation factors in: (a) DY production, (b) DIS, (c) $e^+e^-$ 
annihilation, (d) prompt-photon photoproduciton and (e) prompt-photon 
hadroproduction.}      
\end{figure}

%In the DY hadroproduction (Fig.~\ref{fig})
%of a colourless system (lepton pair, vector boson,
%Higgs boson) of high mass $Q^2$, the vicinity to the threshold region 
In the hadroproduction of a DY lepton pair (Fig.~\ref{fig}a)
of high mass $Q^2$, the vicinity to the threshold region 
is measured by the inelasticity variable $\tau = Q^2/S$, 
where $\sqrt S$ is the centre-of-mass energy. The Born-level
partonic process that controls the cross section is $q{\bar q}$ annihilation. 
In $N$-moment space, where the $N$-moments are defined with respect to $\tau$,
the Sudakov corrections to the $q{\bar q}$-annihilation cross section are taken
into account by a resummation formula analogous to 
Eq.~(\ref{gammaresqq}, \ref{gammaresqg}). Up to NLL accuracy,
the corresponding radiative factor
$\Delta_{DY, \,N}(Q^2)$ has the following explicit expression \cite{Sterman,CT}
%\beq
%\Delta_{DY, \,N}(\as(\mu^2),Q^2/\mu^2;Q^2/\mu_F^2) = 
%\Delta_N^q(\as(\mu^2),Q^2/\mu^2;Q^2/\mu_F^2) \;
%\Delta_N^{\bar q}(\as(\mu^2),Q^2/\mu^2;Q^2/\mu_F^2)
%\eeq
\beq
\label{deltafacdy}
\Delta_{DY, \,N}(Q^2) = \Delta_N^q(Q^2) \;
\Delta_N^{\bar q}(Q^2) \;\;,
\eeq
where $\Delta_N^q(Q^2)$ and $\Delta_N^{\bar q}(Q^2)$
are the single-parton contributions\footnote{To simplify the notation,
we drop the explicit dependence on $\as$ and on the renormalization and
factorization scale. Therefore, we use $\Delta_N^a(Q^2) \equiv 
\Delta_N^q(\as(\mu^2),Q^2/\mu^2;Q^2/\mu_F^2)$ and 
$J_N^a(Q^2) \equiv J_N^a(\as(\mu^2),Q^2/\mu^2)$ throughout this Section.}
given in Eq.~(\ref{deltams}). Each term $\Delta_N^a$ embodies multiple
%the all-order
initial-state radiation 
%from the initial-state parton $a$ 
of soft gluons, i.e. gluons that
carry a small fraction $1-z \sim 1/N \sim (1-\tau)$ of the energy of the
%parton 
initial-state parton $a$. 
%Note that 
The factorized structure on the right-hand
side of Eq.~(\ref{deltafacdy}) implies that soft-gluon interferences between
the two hard partons cancel to this logarithmic accuracy \cite{CMW}. 

Note that this
cancellation does not depend on the type of annihilating partons.
In fact, when the DY pair is replaced by a colourless system, say, a 
Higgs boson, produced by gluon-gluon fusion, the resummed partonic cross
section is controlled by a NLL radiative factor \cite{CdET, Spira} 
\beq
\label{deltafachiggs}
\Delta_{Higgs, \,N}(Q^2) = \Delta_N^g(Q^2) \;
\Delta_N^g(Q^2) \;\;,
\eeq
which is again factorized in single-parton contributions.

The presence of non-interfering Sudakov factors is typical of other
processes dominated by hard scattering of two QCD partons, such as 
lepton-hadron DIS,
%deep-inelastic-scattering (DIS), 
$e^+e^-$ annihilation
in two jets and prompt-photon photoproduction.

In the case of inclusive DIS (Fig.~\ref{fig}b), the hard-scattering scale 
$Q^2= - q^2$ is given by the square of the space-like transferred momentum 
$q$ and the relevant inelasticity variable is the Bjorken variable $x_{Bj}=
Q^2/2P_1 \cdot q$. The Born-level partonic process is lepton-quark scattering 
and, when the threshold region $x_{Bj} \to 1$ is approched, the corresponding
radiative factor $\Delta_{DIS, \,N}(Q^2)$ in $N$-moment space is 
%given by 
\cite{CMW, Sterman96}
%\beq
%\Delta_{DIS, \,N}(\as(\mu^2),Q^2/\mu^2;Q^2/\mu_F^2) = 
%\Delta_N^q(\as(\mu^2),Q^2/\mu^2;Q^2/\mu_F^2) \;
%J_N^q(\as(\mu^2),Q^2/\mu^2)
%\eeq
\beq
\label{deltafacdis}
\Delta_{DIS, \,N}(Q^2) = \Delta_N^q(Q^2) \; J_N^q(Q^2).
\eeq
The Sudakov factor $\Delta_N^q(Q^2)$ is exactly the same as in the DY
process. It embodies soft-gluon radiation from the initial-state quark.
Unlike in the DY process, however, in DIS the scattered initial-state quark
fragments in the final state. Then the factor $J_N^q(Q^2)$ takes into account
the fragmentation of the final-state quark into a jet of collinear and/or soft
partons with a small invariant mass $k^2 \sim Q^2/N \sim (1-x_{Bj})Q^2$.
The NLL expression of the jet mass distribution $J_N^a(Q^2)$ is given in
Eq.~(\ref{jfun}). 

Hadronic final states with two back-to-back jets produced in 
$e^+e^-$ annihilation at the centre-of-mass energy $Q$ (Fig.~\ref{fig}c)
are also controlled
by the jet mass distribution $J_N^a(Q^2)$ \cite{CTTW}. For instance, in the
case of the distribution $(1/\sigma) \,d\sigma/dT$
%$R(T,Q^2) = (1/\sigma) \,d\sigma/dT$
%$R(T,Q^2) = 1/\sigma d\sigma/dT$
%$R(T,Q^2) = d\sigma/\sigma dT$
%$R(T,Q^2) = \frac{d\sigma}{\sigma dT}$
%$\Sigma(T,Q^2)$ 
of the thrust $T$ \cite{thrust},
the Sudakov region is $T \to 1$. In this limit we have 
$1-T \simeq k_1^2/Q^2 + k_2^2/Q^2$, where $k_1^2$ and $k_2^2$ are the hadronic
invariant masses in the two emispheres singled out by the plane orthogonal to
the thrust axis. Considering the $N$-moments 
$\Delta_{T(e^+e^-), \,N}(Q^2)$ of the thrust distribution
%$R(T,Q^2)$ 
with respect to $T$, and taking the large-$N$ limit, one obtains \cite{thrust}
%\beq
%\Delta_{T(e^+e^-), \,N}(\as(\mu^2),Q^2/\mu^2)
%= J_N^q(\as(\mu^2),Q^2/\mu^2) \;
%J_N^{\bar q}(\as(\mu^2),Q^2/\mu^2)  
%\eeq
\beq
\label{deltafacT}
\Delta_{T(e^+e^-), \,N}(Q^2) = J_N^q(Q^2) \;J_N^{\bar q}(Q^2) \;\;.  
\eeq
The factors $J^q$ and $J^{\bar q}$ are the invariant-mass distributions
of the two jets that originate from the fragmentation of the 
$q{\bar q}$-pair produced by the $e^+e^-$-annihilation process at the Born
level.

The structure of the radiative factors in 
Eqs.~(\ref{deltafacdy}--\ref{deltafacT}) easily explains the high-$E_T$
behaviour of the prompt-photon cross section  in
photoproduction collisions (Fig.~\ref{fig}d). This process, which can be regarded 
as a simplified case of the hadroproduction process considered throughout the
paper, is discussed in Appendix~C. In hadron-photon collisions the production
of high-$E_T$ prompt photons is dominated at the Born level by the 
Compton-scattering subprocess  $q(x_1P_1) + \gamma(p_2) \to q(p_3) + \gamma(p)$.
The all-order resummation of 
%soft-gluon 
Sudakov effects leads to the radiative factor
$\Delta_N^{q\gamma \to q \gamma}$ in Eq.~(\ref{photresqg}),
whose NLL expression is given in Eq.~(\ref{deltaphot}):
%\beq
%\Delta_N^{q\gamma \to q \gamma}(\as(\mu^2),Q^2/\mu^2;Q^2/\mu_F^2) =
%\Delta_N^q(\as(\mu^2),Q^2/\mu^2;Q^2/\mu_F^2) \;
%J_N^q(\as(\mu^2),Q^2/\mu^2)
%\eeq
\beq
\label{deltafacphot}
\Delta_N^{q\gamma \to q \gamma}(Q^2) = \Delta_N^q(Q^2) \;
J_N^q(Q^2) \;\;.
\eeq
The factor $\Delta_N^q(Q^2)$ takes into account sof-gluon radiation from
the initial-state quark, while $J_N^q(Q^2)$ is the mass distribution of the
jet produced by the collinear and/or soft fragmentation of the final-state
quark.

Note that high-$E_T$
prompt-photon photoproduction can be regarded as a 
%`non point-like'
photon-hadron deep-inelastic scattering, 
% of the incoming hadron from  the  
where the space-like momentum transferred by the scattered photon is 
$q^\mu = p_2^\mu - p^\mu$. 
Since the high-$E_T$
%prompt-photon 
cross section is dominated by the kinematics configurations in which the 
prompt photon is produced in the central rapidity region, we 
%also 
have 
$2p_2\cdot p \simeq 2E_T^2$ and $2P_1 \cdot p \simeq P_1 \cdot p_2 = S/2$.
Thus, the hard scale is $Q^2 = - q^2= 2p_2\cdot p \simeq 2E_T^2$
and the inelasticity variable analogous to the Bjorken variable is
$Q^2/2P_1 \cdot q = Q^2 /(2P_1 \cdot p_2 - 2P_1 \cdot p) \simeq 4E_T^2/S =
x_T^2$. Recalling that in Eq.~(\ref{deltafacphot}) we have $Q^2 = 2E_T^2$
(see Eq.~(\ref{etscalephot}))
and that $N$ is the moment index with respect to $x_T^2$
(see Eq.~(\ref{shnphot})), we can thus straigtforwardly understand the complete
analogy between Eq.~(\ref{deltafacphot}) and the expression
(\ref{deltafacdis}) for the DIS radiative factor.

The Sudakov corrections to prompt-photon hadroproduction (Fig.~\ref{fig}e)
are embodied in
Eqs.~(\ref{gammaresqq}, \ref{gammaresqg})
through the radiative factor $\Delta_N^{ab \to d \gamma}(Q^2)$.
On the basis of the factorization of the right-hand side of
Eqs.~(\ref{deltafacdy}--\ref{deltafacphot})
in terms of initial- and final-state single-parton contributions,
one might expect that $\Delta_N^{ab \to d \gamma}$ can be obtained from
the photoproduction result in Eq.~(\ref{deltafacphot}) by simply including
an additional initial-state factor $\Delta_N^b$. The NLL expression 
(\ref{deltangamma}) for $\Delta_N^{ab \to d \gamma}(Q^2)$ shows that this
naive expectation is not correct. In fact, we have
%\beeq
%\label{deltafachad}
%\Delta_N^{ab \to d \gamma}(\as(\mu^2),Q^2/\mu^2;Q^2/\mu_F^2) &=&
%\Delta_N^a(\as(\mu^2),Q^2/\mu^2;Q^2/\mu_F^2)
%\; \Delta_N^b(\as(\mu^2),Q^2/\mu^2;Q^2/\mu_F^2) 
%\nonumber \\
%&\cdot& J_N^d(\as(\mu^2),Q^2/\mu^2) \;
%\Delta_N^{({\rm int}) \,ab \to d \gamma}(\as(\mu^2),Q^2/\mu^2) \;\;.
%\eeeq
\beq
\label{deltafachad}
\Delta_N^{ab \to d \gamma}(Q^2) =
\Delta_N^a(Q^2) \; \Delta_N^b(Q^2) 
\;J_N^d(Q^2) \;\Delta_N^{({\rm int}) \,ab \to d \gamma}(Q^2) \;\;.
\eeq

The presence of the NLL contribution $\Delta_N^{({\rm int}) \,ab \to d \gamma}$
on the right-hand side of Eq.~(\ref{deltafachad})
implies that the physical picture of the Sudakov
%corrections 
radiative factors
in terms of {\em independent} single-parton contributions is not 
valid, in general. As discussed in Sect.~\ref{secgen} and explicitly shown in 
Eqs.~(\ref{deltafacdy}--\ref{deltafacphot}), this picture applies to
processes dominated by hard-scattering of two sole partons, but it breaks down
at NLL accuracy in the case of multiparton hard-scattering. The breakdown 
is due to {\em interferences} and {\em colour correlations} produced by soft 
gluons that are radiated at large angle with respect to the directions of the 
hard-parton momenta \cite{inprep}. Soft-gluon interferences are present in the  
hard-scattering of three QCD partons as shown by Eq.~(\ref{deltafachad}),
while colour correlations affect hard-scattering of more than three QCD partons
\cite{kidon}.

Owing to their large-angle origin, soft-gluon interferences are process
dependent. In the case of prompt-photon hadroproduction they are taken into
account by the factor $\Delta_N^{({\rm int}) \,ab \to d \gamma}$, whose
explicit NLL expression is given in Eqs.~(\ref{deltaintgamma}--\ref{D1coef}).

Note that the coefficient $D_{ab \to d \gamma}^{(1)}$ in
Eq.~(\ref{D1coef}) depends linearly on the colour charges of the hard partons,
and the colour-charge dependence of 
$\Delta_N^{({\rm int}) \,ab \to d \gamma}$ is thus factorized at NLL accuracy.
This suggests that the effect of the interference factor can be absorbed by
a proper rescaling of the independent-emission factors $\Delta^a, \Delta^b$
and $J^d$. As a matter of fact, neglecting 
corrections beyond NLL order, one can check that the right-hand side of
Eq.~(\ref{deltafachad}) can be rewritten as follows
\beq
\label{deltafachoer}
\Delta_N^{ab \to d \gamma}(Q^2) =
\Delta_{N/2}^a(Q^2/2) \; \Delta_{N/2}^b(Q^2/2) 
\;J_{N/2}^d(Q^2/2)  \;\;.
\eeq
This equation has to be regarded as a manifestation of the 
colour-coherence properties of QCD emission \cite{BCM}.
Soft gluons radiated at large angle 
%with respect to the directions of the hard-parton momenta 
destructively interfere. Their effect can thus be taken into account by
Sudakov factors of independent emission in a 
restricted (angular) region of the phase space.
\section{Conclusion}
\label{secfin}
In this paper we presented the explicit expressions for the resummation of
threshold-enhanced logarithms in hadronic prompt-photon production,
to next-to-leading accuracy. 
The simple colour structure of the diagrams
contributing to prompt-photon production reflects itself in the simplicity of
the resummed formulae. Fragmentation processes, furthermore, do
not contribute to the Sudakov resummation at NLL level.
In Mellin space, the resummed radiative factor
factorizes in the product of three independent contributions for the initial
and  final coloured partons appearing in the Born process, times a simple
factor describing the soft-gluon interferences between initial and final
states.  General coherence properties of large-angle soft-gluon radiation allow
to further simplify the result: the interference contributions can be
described, to the same degree of accuracy, by constraining the phase-space for
independent emission from the coloured partons. The resulting radiation factor
can thus be written as the product of the three independent single-parton           
contributions, with a properly rescaled dependence on the Mellin-moment
variable $N$.

The formulae are given in terms of
Mellin moments, and can be used for 
phenomenological applications by inverse-Mellin transforming to $x_T$ space. The
problems related to this inversion are the same as those encountered in the
resummation of the Drell-Yan or heavy-quark production cross-sections, and can
therefore be solved with the same techniques~\cite{CMNT2}. All ingredients are
therefore available for a phenomenological study of prompt-photon production
including the evaluation of Sudakov effects with NLL accuracy. Such a study is
in progress, and will be reported soon. The calculations presented in this
work, together with previous work on Drell-Yan and DIS, make it now possible
to carry out global fits of parton densities with a uniform NLL accuracy
in the large-$x$ region. All of the processes that are used for these global
fits, among which prompt-photon production plays a critical role, are
now known theoretically at this level of accuracy.                   
\\[1cm]
{\bf Acknowledgements} We thank W. Vogelsang for useful discussions.
\\[2cm]
%\section{Appendix A: NLL formulae for the radiative factors}
%\label{secnllform}

\noindent {\bf \Large Appendix A: NLL formulae for the radiative factors}
\bigskip

\noindent The logarithmic expansion of the 
radiative factors in Eqs.~(\ref{deltams}, \ref{jfun}, \ref{deltaintgamma})
can be computed as described in Refs.~\cite{CT, CTTW}. The running coupling
$\as(k^2)$ with $k^2=q^2, (1-z)^2Q^2, (1-z)Q^2$ has to be expressed in terms
of $\as(\mu^2)$ according to the NLO solution of the renormalization group
equation:
\beeq
\label{rgesol}
\as(k^2) &=& \frac{\as(\mu^2)}{1+ b_0 \as(\mu^2) \ln (k^2/\mu^2)} 
\Bigl[ 1 - \frac{b_1}{b_0}
\frac{\as(\mu^2)}{1+ b_0 \as(\mu^2) \ln (k^2/\mu^2)} 
\ln (1+ b_0 \as(\mu^2) \ln (k^2/\mu^2))
\nonumber \\
&+& {\cal O}(\as^2(\mu^2) (\as(\mu^2) \ln (k^2/\mu^2))^n) \Bigr] \;, 
\eeeq
where $b_0, b_1$ are the first two coefficients of the QCD $\beta$-function,
which are explicitly reported in Eq.~(\ref{betass}).
Then the $z$ integration
can be performed with NLL accuracy by setting
\beq
\label{expfac}
z^{N-1}-1 \simeq -\Theta(1-z - e^{-\GE}/N ) \;\;.
\eeq

Defining
\beq
\label{lamdef}
\lambda = b_0 \as(\mu^2) \ln N \;\;,
\eeq
we find
\beeq
\label{lndeltams}
\!\!\! \!\!\! \!\!\! \!\!\! \!\!\!
\ln \Delta_N^a(\as(\mu^2),Q^2/\mu^2;Q^2/\mu_F^2) 
&\!\!=\!\!& \ln N \;h_a^{(1)}(\lambda) +
h_a^{(2)}(\lambda,Q^2/\mu^2;Q^2/\mu_F^2) + 
{\cal O}\left(\as(\as \ln N)^k\right) \,,\\
%\eeq
%\beq
\label{lnjfun}
\ln J_N^a(\as(\mu^2),Q^2/\mu^2) &\!\!=\!\!& \ln N \;f_a^{(1)}(\lambda) +
f_a^{(2)}(\lambda,Q^2/\mu^2) + {\cal O}\left(\as(\as \ln N)^k\right) \,,\\
%\eeq
%\beq
\label{lndeltaintgamma}
\ln \Delta_N^{({\rm int}) \,ab \to d \gamma}(\as(\mu^2),Q^2/\mu^2)
&\!\!=\!\!& \frac{D_{ab \to d \gamma}^{(1)}}{2\pi b_0} \;\ln (1-2\lambda) +
{\cal O}\left(\as(\as \ln N)^k\right) \,,
\eeeq
where the LL and NLL functions $h_a^{(1)},f_a^{(1)}$ and
$h_a^{(2)},f_a^{(2)}$ are given in terms of the perturbative coefficients
$A_a^{(1)}, A_a^{(2)}, B_a^{(1)}$ in Eqs.~(\ref{Afun}, \ref{Bfun}):
\beeq
\label{hll}
h_a^{(1)}(\lambda) =
&+&\frac{A_a^{(1)}}{2\pi b_0 \lambda}\Bigl[ 2\lambda + (1-2\lambda)
\ln(1-2\lambda) \Bigr] \; , \\
\label{hnll}
h_a^{(2)}(\lambda,Q^2/\mu^2;Q^2/\mu_F^2) =
&+&\frac{A_a^{(1)} b_1}{2\pi b_0^3}\Bigl[ 2\lambda + \ln(1-2\lambda)
+\half\ln^2(1-2\lambda) \Bigr] \nonumber \\
&-& \frac{A_a^{(1)}\GE}{\pi b_0} \ln(1-2\lambda) \\
&-&\frac{A_a^{(2)}}{2\pi^2 b_0^2}\Bigl[ 2\lambda + \ln(1-2\lambda)\Bigr] 
+ \frac{A_a^{(1)}}{2\pi b_0}\Bigl[ 2\lambda + \ln(1-2\lambda)\Bigr]
\ln\frac{Q^2}{\mu^2}
- \frac{A_a^{(1)}}{\pi b_0} \lambda \ln\frac{Q^2}{\mu_F^2}
\; , \nonumber
\eeeq
\beeq
\label{fll}
f_a^{(1)}(\lambda) =
&-&\frac{A_a^{(1)}}{2\pi b_0 \lambda}\Bigl[(1-2\lambda)
\ln(1-2\lambda)-2(1-\lambda)
\ln(1-\lambda)\Bigr] \; , \\
\label{fnll}
f_a^{(2)}(\lambda,Q^2/\mu^2) =
&-&\frac{A_a^{(1)} b_1}{2\pi b_0^3}\Bigl[\ln(1-2\lambda)
-2\ln(1-\lambda)+\half\ln^2(1-2\lambda)-\ln^2(1-\lambda)\Bigr] \nonumber \\
&+&\frac{B_a^{(1)}}{2\pi b_0}\ln(1-\lambda)
-\frac{A_a^{(1)}\GE}{\pi b_0}\Bigl[\ln(1-\lambda)
-\ln(1-2\lambda)\Bigr] \\
&-&\frac{A_a^{(2)}}{2\pi^2 b_0^2}\Bigl[2\ln(1-\lambda)
-\ln(1-2\lambda)\Bigr] 
+ \frac{A_a^{(1)}}{2\pi b_0}\Bigl[2\ln(1-\lambda)
-\ln(1-2\lambda)\Bigr] \ln\frac{Q^2}{\mu^2} \; . \nonumber
\eeeq

Note that the functions $f_a^{(1)}(\lambda)$ and $f_a^{(2)}(\lambda,Q^2/\mu^2)$
can also be written in terms of $h_a^{(1)}$ and 
$h_a^{(2)}$ as follows
\beeq
\label{fllh}
f_a^{(1)}(\lambda) &=& h_a^{(1)}(\lambda/2)- h_a^{(1)}(\lambda) \;, \\
%\eeq
%\beq
\label{fnllh}
f_a^{(2)}(\lambda,Q^2/\mu^2) &=& 2 \,h_a^{(2)}(\lambda/2, Q^2/\mu^2;1)- 
h_a^{(2)}(\lambda, Q^2/\mu^2;1) 
+\frac{B_a^{(1)}+ 2A_a^{(1)}\GE}{2\pi b_0}\ln(1-\lambda) \;.
\eeeq

Inserting the expressions 
(\ref{lndeltams}, \ref{lnjfun}, \ref{lndeltaintgamma}) 
into Eq.~(\ref{deltangamma}), and using the explicit form of the perturbative
coefficients $A_a^{(1)}, A_a^{(2)}, B_a^{(1)}, D_{ab \to d \gamma}^{(1)}$
in Eqs.~(\ref{A12coef}, \ref{B1coef}, \ref{D1coef}) we obtain the results
in Eqs.~(\ref{deltanll}--\ref{g2qg}).

\bigskip

\bigskip

\noindent {\bf \Large Appendix B: NNLO partonic cross sections at large $N$}
\bigskip

\noindent According to the notation in Eq.~(\ref{sigtow}), the large-$N$
behaviour of the NLO 
%and NNLO 
cross section
${\hat \sigma}_{ab\to \gamma}^{(1)}$ 
%and ${\hat \sigma}_{ab\to \gamma}^{(2)}$
in Eq.~(\ref{pxsg}) is written as
\beeq
{\hat \sigma}_{ab\to \gamma, \,N}^{(1)}(E_T^2, \mu^2, \mu_F^2, \mu_f^2) &=&
{\hat \sigma}_{ab\to d\gamma, \,N}^{(0)} \;
\Bigl[ c_{1,2}^{(ab)} \;\ln^2 N + c_{1,1}^{(ab)}(E_T^2/\mu_F^2) \ln N
+ c_{1,0}^{(ab)}(E_T^2/\mu_F^2,E_T^2/\mu^2) \Bigr. \nonumber \\
\label{sig1N}
&+& \Bigl. {\cal O}(1/N) \Bigr] \;\;,
\eeeq
where the various coefficients can be read from Eqs.~(\ref{qq1st}, \ref{qg1st})
\beq
\label{c12coef}
c_{1,2}^{(q{\bar q})} = \frac{1}{\pi} \left( 2C_F - \frac{1}{2} C_A \right)
\;\;, \quad 
c_{1,2}^{(qg)} = \frac{1}{\pi} \left( \frac{1}{2} C_F + C_A \right) \;\;,
\eeq
\beeq
\label{c11qqcoef}
c_{1,1}^{(q{\bar q})}(E_T^2/\mu_F^2) &=& \frac{1}{\pi}
\left[ \GE (4 C_F - C_A) - (2 C_F - C_A) \ln 2
+ \pi b_0 - 2 C_F \ln \frac{2E_T^2}{\mu_F^2} \;\right] \;\;, \nonumber \\
\label{c11qgcoef}
c_{1,1}^{(qg)}(E_T^2/\mu_F^2) &=& \frac{1}{\pi}
\left[ \GE ( C_F + 2 C_A) - C_A \ln 2
+ \frac{3}{4} C_F - ( C_F + C_A ) \ln \frac{2E_T^2}{\mu_F^2} \;\right] \;\;,
\eeeq
\beeq
c_{1,0}^{(q{\bar q})}(E_T^2/\mu_F^2,E_T^2/\mu^2) &=& \frac{1}{\pi}
C_{q{\bar q}\to \gamma}^{(1)}(2E_T^2/\mu^2;2E_T^2/\mu_F^2) \;\;, 
\nonumber \\
%\quad
\label{c10qgcoef}
c_{1,0}^{(qg)}(E_T^2/\mu_F^2,E_T^2/\mu^2) &=& \frac{1}{\pi}
C_{qg\to \gamma}^{(1)}(2E_T^2/\mu^2;2E_T^2/\mu_F^2) \;\;,
\eeeq
and $C_{q{\bar q}\to \gamma}^{(1)}, C_{qg\to \gamma}^{(1)}$ are given in 
Eqs.~(\ref{cqq1coef}, \ref{cqg1coef}).

Analogously, we can write the NNLO cross section
${\hat \sigma}_{ab\to \gamma}^{(2)}$ as follows:
\beeq
{\hat \sigma}_{ab\to \gamma, \,N}^{(2)}(E_T^2, \mu^2, \mu_F^2, \mu_f^2) &=&
{\hat \sigma}_{ab\to d\gamma, \,N}^{(0)} \;
\Bigl[ c_{2,4}^{(ab)} \;\ln^4 N + c_{2,3}^{(ab)}(E_T^2/\mu_F^2) \ln^3 N
\Bigr. \nonumber \\
\label{sig2N} 
&+& \Bigl. c_{2,2}^{(ab)}(E_T^2/\mu_F^2,E_T^2/\mu^2) \ln^2 N
+ {\cal O}(\ln N) \Bigr] \;\;.
\eeeq
The coefficients $c_{2,4}, c_{2,3}, c_{2,2}$ can be calculated by expanding
the resummation formulae (\ref{gammaresqq}, \ref{gammaresqg}) to the
second order in $\as$. We find
\beq
\label{c24coef}
c_{2,4}^{(ab)} = \frac{1}{2} \;\left[ c_{1,2}^{(ab)} \right]^2 \;\;,
\eeq
\beeq
c_{2,3}^{(q{\bar q})}(E_T^2/\mu_F^2) &=& 
c_{1,2}^{(q{\bar q})} \, c_{1,1}^{(q{\bar q})}(E_T^2/\mu_F^2) 
+ \frac{2}{3\pi} b_0 \left( 2 C_F - \frac{3}{4} C_A \right) \;\;,
\nonumber \\
\label{c23qgcoef}
c_{2,3}^{(qg)}(E_T^2/\mu_F^2) &=& 
c_{1,2}^{(qg)} \, c_{1,1}^{(qg)}(E_T^2/\mu_F^2)
+ \frac{2}{3\pi} b_0 \left( \frac{1}{4} C_F + C_A \right) \;\;,
\eeeq
\beeq
&&c_{2,2}^{(q{\bar q})}(E_T^2/\mu_F^2,E_T^2/\mu^2) =
\frac{1}{2} \;\left[ c_{1,1}^{(q{\bar q})}(E_T^2/\mu_F^2) \right]^2
+ c_{1,2}^{(q{\bar q})} \;c_{1,0}^{(q{\bar q})}(E_T^2/\mu_F^2,E_T^2/\mu^2)
\nonumber \\
&&\;\; \;\;\;\;+ \frac{1}{\pi} b_0 \left[ \GE \Bigl( 4 C_F - \frac{3}{2} 
C_A \Bigr)
- (2 C_F - C_A) \ln 2 + \frac{1}{2} \pi b_0 + 
\Bigl( 2 C_F - \frac{1}{2} C_A \Bigr) 
\left( \frac{K}{2 \pi b_0} - \ln \frac{2 E_T^2}{\mu^2} \right)
\right] \;\;,
\nonumber \\
&&c_{2,2}^{(qg)}(E_T^2/\mu_F^2,E_T^2/\mu^2) =
\frac{1}{2} \;\left[ c_{1,1}^{(qg)}(E_T^2/\mu_F^2) \right]^2
+ c_{1,2}^{(qg)} \;c_{1,0}^{(qg)}(E_T^2/\mu_F^2,E_T^2/\mu^2)
\nonumber \\
&&\;\; \;\;\;\;+ \frac{1}{\pi} b_0 \left[ \GE \Bigl( \frac{1}{2} C_F + 2 C_A 
\Bigr)
- C_A \ln 2 + \frac{3}{8} C_F +
\Bigl( \frac{1}{2} C_F +  C_A \Bigr)
\left( \frac{K}{2 \pi b_0} - \ln \frac{2 E_T^2}{\mu^2} \right)
\right] \;\;, 
\eeeq
where the coefficients $c_{1,2}, \,c_{1,1}, \,c_{1,0}$ and $K$ are given in
Eqs.~(\ref{c12coef}, \ref{c11qgcoef}, \ref{c10qgcoef}) and (\ref{kcoef}).

Our prediction for the coefficients in Eq.~(\ref{sig2N}) can be used to check
future NNLO calculations of the prompt-photon production cross section.
Alternatively, when these calculations become available, they can provide
a highly non-trivial check of our NNL resummation. 

\newpage
 
%\section{Appendix C: photoproduction of prompt photons}
%\label{secphotprod}

\bigskip

\bigskip
\noindent {\bf \Large Appendix C: photoproduction of prompt photons}
\bigskip

\noindent In hadron-photon collisions the inclusive production of a single
prompt photon is due to the process
\beq
\label{photgamma}
H_1(P_1) + \gamma(P_2) \to \gamma(p) + X \;\;.
\eeq
We use the same kinematics notation as in the hadroproduction case (cf.
Sect.~\ref{secnotat}) and we write the prompt-photon photoproduction 
cross section integrated over $\eta$ at fixed $E_T$ as follows:
\beq
\label{1pxsgammaphot}
\frac{d\sigma_{\gamma}^{(\rm ph)}(x_T,E_T)}{d E_T} =
\left( \frac{d\sigma_{\gamma}^{(\rm ph)}(x_T,E_T)}{d E_T} 
\right)_{\!\rm hadronic} + \;
\left( \frac{d\sigma_{\gamma}^{(\rm ph)}(x_T,E_T)}{d E_T}
\right)_{\!\rm pointlike} \;\;.
\eeq
The hadronic contribution to the cross section is completely analogous to
the right-hand side of Eq.~(\ref{1pxsgamma}) apart from replacing 
$f_{b/H_2}(x_2,\mu_F^2)$ with the parton distribution
$f_{b/\gamma}(x_2,\mu_F^2)$ of incoming photon.

The second contribution on the right-hand side of Eq.~(\ref{1pxsgammaphot})
is due to point-like interactions of the incoming photon with high-momentum
partons. The point-like cross section can in turn be decomposed in direct
and fragmentation components
\beeq
\label{1pxsgammapl}
\left( \frac{d\sigma_{\gamma}^{(\rm ph)}(x_T,E_T)}{d E_T}
\right)_{\!\!\rm pointlike}
&\!\!\!\!\!=&\!\! \frac{1}{E_T^3} \sum_{a}
\int_0^1 dx_1 \;f_{a/H_1}(x_1,\mu_F^2)  \nonumber \\
&\!\!\!\!\!\cdot&\!\! \int_0^1 dx \left\{
\delta\!\left(x - \frac{x_T}{{\sqrt {x_1}}} \right) 
{\hat \sigma}_{a\gamma\to {\gamma}}(x, \as(\mu^2); E_T^2, 
\mu^2, \mu_F^2, \mu_f^2) \!\! \!\! \right. \!\! \!\!\!\! \\
&\!\!\!\!\!+&\!\!\! \left. \sum_{c} \int_0^1 dz \;z^2 \;d_{c/\gamma}(z,\mu_f^2)
\;\delta\!\left(x - \frac{x_T}{z{\sqrt {x_1}}} \right)
{\hat \sigma}_{a\gamma\to c}(x, \as(\mu^2); E_T^2, \mu^2, \mu_F^2, \mu_f^2)
\right\} . \nonumber
\eeeq
The rescaled partonic cross sections ${\hat \sigma}_{a\gamma\to \gamma}$
and ${\hat \sigma}_{a\gamma\to c}$ have perturbative QCD expansions
similar to Eqs.~(\ref{pxsg}) and (\ref{pxsc}). In particular, for the
point-like direct component we have 
\beq
\label{pxsgphot}
{\hat \sigma}_{a\gamma\to \gamma}(x, \as(\mu^2); E_T^2, \mu^2, \mu_F^2, \mu_f^2)
= \alpha^2 \, \left[
{\hat \sigma}_{a\gamma\to d \gamma}^{(0)}(x) +
\sum_{n=1}^{\infty} \as^n(\mu^2) \, 
{\hat \sigma}_{a\gamma\to \gamma}^{(n)}(x; E_T^2, \mu^2, \mu_F^2, \mu_f^2)
\right] \;,
\eeq
where the only non-vanishing terms at LO are those due
to the Compton scattering subprocesses 
\beq
\label{parprophot}
q + \gamma \to q + \gamma \;\;, \quad \quad 
{\bar q} + \gamma \to {\bar q} + \gamma \;\;,
\eeq
%and the corresponding 
whose contribution to the cross section is
\beq
\label{siqgamma}
{\hat \sigma}_{q\gamma \to q\gamma}^{(0)}(x) = 
{\hat \sigma}_{{\bar q}\gamma \to {\bar q}\gamma}^{(0)}(x) = \pi \,
e_q^4 \;\frac{x^2}{\sqrt {1-x^2}}
\left(1 + \frac{x^2}{4}\right) \;\;. 
\eeq

To perform soft-gluon resummation at high $E_T$, we work as usual in 
$N$-moment space by defining
\beq
\label{shnphot}
\sigma_{\gamma, \,N}^{(\rm ph)}(E_T) \equiv \int_0^1 dx_T^2 \;(x_T^2)^{N-1} 
\;E_T^3 \frac{d\sigma_\gamma^{(\rm ph)}(x_T,E_T)}{d E_T} \;\;.
\eeq 

The resummation of the large-$N$ corrections to the $N$-moments
of the hadronic contribution in Eq.~(\ref{1pxsgammaphot}) is exactly the same 
as for the hadroproduction case discussed in Sect.~\ref{secresum}.
Moreover, in the large-$N$ limit, the point-like contribution turns out to be
dominant: the hadronic contribution involves the 
additional convolution with the photon parton density $f_{b/\gamma}$ and
this implies its suppression by a relative factor of ${\cal O}(1/N)$. We can
thus limit ourselves to considering the point-like cross section.

In the case of the point-like contribution, one can repeat the argument in 
Sect.~\ref{secnll} on the relative size of the fragmentation component and of
the various direct subprocesses. Up to NLL accuracy, we then conclude that 
soft-gluon resummation in the photoproduction cross section 
(\ref{shnphot}) is controlled by the point-like direct channels
$q\gamma \to \gamma$ and ${\bar q}\gamma \to \gamma$. The 
all-order resummation formulae for the corresponding partonic cross sections
are
\beeq
%\label{gammaresqg}
{\hat \sigma}_{q\gamma\to \gamma, \;N}^{({\rm res})}(\as(\mu^2); 
E_T^2, \mu^2, \mu_F^2, \mu_f^2) &=& \alpha^2  
\;{\hat \sigma}_{q\gamma\to q\gamma, \;N}^{(0)} 
\;C_{q\gamma \to \gamma}(\as(\mu^2),Q^2/\mu^2;Q^2/\mu_F^2) \nonumber \\
\label{photresqg}
&\cdot& 
\Delta_{N+1}^{q\gamma \to q \gamma}(\as(\mu^2),Q^2/\mu^2;Q^2/\mu_F^2)
\;\;, \\
%\eeeq
%\beeq
\label{photresqbarg}
{\hat \sigma}_{{\bar q}\gamma\to \gamma, \;N}^{({\rm res})}(\as(\mu^2); 
E_T^2, \mu^2, \mu_F^2, \mu_f^2) &=& 
{\hat \sigma}_{q\gamma\to \gamma, \;N}^{({\rm res})}(\as(\mu^2); 
E_T^2, \mu^2, \mu_F^2, \mu_f^2)
\;\;,
\eeeq 
where 
%$Q^2 = 2 E_T^2$.
\beq
\label{etscalephot}
Q^2 = 2 E_T^2 \;\;,
\eeq
and ${\hat \sigma}_{q\gamma\to q\gamma, \;N}^{(0)}$ are the $N$-moments
with respect to $x^2$ of Eq.~(\ref{siqgamma})
\beq
\label{photgammaN}
{\hat \sigma}_{q\gamma\to q\gamma, \;N}^{(0)} = \pi \,e_q^2 
\,\frac{1}{4} \;\frac{\Gamma(1/2) \;\Gamma(N+1)}{\Gamma(N+5/2)}
\;(7+5N) \;\;.
\eeq
The radiative factor $\Delta_N^{q\gamma \to q \gamma}$ and the
$N$-independent function $C_{q\gamma \to \gamma}$ in Eq.~(\ref{gammaresqg})
can directly be related to the analogous contributions
$\Delta_N^{qg \to q \gamma}$ and $C_{qg \to \gamma}$ to the $qg$ channel
in the hadroproduction process.

The radiative factor $\Delta_N^{q\gamma \to q \gamma}$ is obtained from
the factorized expression (\ref{deltangamma}) for $\Delta_N^{qg \to q \gamma}$,
namely from $\Delta_N^{qg \to q \gamma} = \Delta_N^q \Delta_N^g J_N^q
\Delta_N^{({\rm int}) \,qg \to q \gamma}$, by switching off soft-gluon
radiation from the incoming gluon. This amounts to set $C_A=0$ in $\Delta_N^g$
and $\Delta_N^{({\rm int}) \,qg \to q \gamma}$. Using the explicit formulae
in Eqs.~(\ref{deltams}, \ref{A12coef}) and 
(\ref{deltaintgamma}, \ref{D1coef}), this implies
that up to NLL accuracy we can neglect both $\Delta_N^g$ and 
$\Delta_N^{({\rm int}) \,qg \to q \gamma}$ and we have the simple result:
\beq
\label{deltaphot}
\Delta_N^{q\gamma \to q \gamma}(\as(\mu^2),Q^2/\mu^2;Q^2/\mu_F^2) 
= \Delta_N^q(\as(\mu^2),Q^2/\mu^2;Q^2/\mu_F^2) \;
J_N^q(\as(\mu^2),Q^2/\mu^2) \;\;.
\eeq

Note that no soft-gluon interference factor 
$\Delta_N^{({\rm int})}$ appears in Eq.~(\ref{deltaphot}).
Prompt-photon photoproduction at threshold is dominated by an
underlying hard-scattering that involves only two hard partons and then,
in agreement with the general discussion in Sect.~\ref{secgen}, soft-gluon
interferences have to cancel.

The explicit NLL expansion of Eq.~(\ref{deltaphot}) gives
\beeq
\Delta_N^{q \gamma \to q \gamma}\!\left(\as(\mu^2),\frac{Q^2}{\mu^2};
\frac{Q^2}{\mu_F^2}\right) &=&
\exp \left\{ \ln N \; g_{q \gamma}^{(1)}(b_0\as(\mu^2)\ln N) +
g_{q \gamma}^{(2)}(b_0\as(\mu^2)\ln N,Q^2/\mu^2;Q^2/\mu_F^2 ) \right. \nonumber \\
\label{deltanllphot}
&+& \left. {\cal O}(\as(\as \ln N)^k) \right\} \;\;;
\eeeq
where the LL and NLL terms $g^{(1)}$ and $g^{(2)}$ are expressed 
in terms of the auxiliary functions $h^{(1)}$ and $h^{(2)}$ of 
Eqs.~(\ref{htll}) and (\ref{htnll}):
\beq
\label{g1funphot}
g_{q\gamma}^{(1)}(\lambda) = C_F \;h^{(1)}(\lambda/2) \;, 
\eeq
\beeq
g_{q\gamma}^{(2)}\!\left(\lambda,\frac{Q^2}{\mu^2};\frac{Q^2}{\mu_F^2}\right)
&=& 2 \,C_F \;h^{(2)}(\lambda/2) 
+ \frac{4 C_F \,\GE - 3 C_F}{4\pi b_0} \ln(1-\lambda) \nonumber \\
\label{g2funphot}
&-&  \frac{C_F}{\pi b_0} \;\lambda \ln \frac{Q^2}{\mu_F^2}
+  \frac{C_F}{\pi b_0} \Bigl[ \lambda    
+  \ln(1-\lambda)  \Bigr]
\; \ln \frac{Q^2}{\mu^2} \;. 
\eeeq

The $N$-independent function $C_{q\gamma\to \gamma}(\as)$ has the 
following perturbative expansion
\beq
\label{cgammaphot}
C_{q\gamma\to \gamma}(\as(\mu^2),Q^2/\mu^2;Q^2/\mu_F^2) = 
1 + \frac{\as(\mu^2)}{\pi} \;C_{q\gamma\to \gamma}^{(1)}(Q^2/\mu_F^2) +
\sum_{n=2}^{+\infty} \; 
\left( \frac{\as(\mu^2)}{\pi} \right)^n 
C_{q\gamma\to \gamma}^{(n)}(Q^2/\mu^2;Q^2/\mu_F^2) \;.
\eeq
Note that the first-order coefficient $C_{q\gamma\to \gamma}^{(1)}$ does not
depend on the renormalization scale. Its explicit expression is obtained 
from that of $C_{qg\to \gamma}^{(1)}$ by
setting $C_A=0$ and $b_0=0$ in Eq.~(\ref{cqg1coef}):
\beq
\label{c1coefphot}
C_{q\gamma\to \gamma}^{(1)}(Q^2/\mu_F^2) = C_F \left\{
\frac{1}{2} \GE^2  + \frac{3}{4} \GE  
- \, \frac{1}{10} \ln 2 - \frac{1}{2} \,\frac{K_q}{C_F} 
+ \frac{\pi^2}{30} + \frac{1}{2}  \ln^2 2
-  \Bigl( \GE  - \frac{3}{4}  \Bigr) 
\ln \frac{Q^2}{\mu_F^2} \right\}
 \;\;. 
\eeq

Expanding the resummation formula (\ref{photresqg}) in powers of $\as$
we can derive the large-$N$ behaviour of the NLO and NNLO 
cross sections ${\hat \sigma}_{q\gamma\to \gamma}^{(1)}$ and
${\hat \sigma}_{q\gamma\to \gamma}^{(2)}$ of Eq.~(\ref{pxsgphot}).

At NLO we find
%\beeq
%{\hat \sigma}_{q\gamma\to \gamma, \,N}^{(1)}(E_T^2, \mu^2, \mu_F^2, \mu_f^2) &=&
%{\hat \sigma}_{q\gamma\to q\gamma, \,N}^{(0)} \;
%\Bigl[ c_{1,2}^{(q\gamma)} \;\ln^2 N + c_{1,1}^{(q\gamma)}(E_T^2/\mu_F^2) \ln N
%+ c_{1,0}^{(q\gamma)}(E_T^2/\mu_F^2) \Bigr. \nonumber \\
%\label{sig1Nphot}
%&+& \Bigl. {\cal O}(1/N) \Bigr] \;\;,
%\eeeq
\beq
\label{sig1Nphot}
{\hat \sigma}_{q\gamma\to \gamma, \,N}^{(1)}(E_T^2, \mu^2, \mu_F^2, \mu_f^2) =
{\hat \sigma}_{q\gamma\to q\gamma, \,N}^{(0)} \;
\Bigl[ c_{1,2}^{(q\gamma)} \;\ln^2 N + c_{1,1}^{(q\gamma)}(E_T^2/\mu_F^2) \ln N
+ c_{1,0}^{(q\gamma)}(E_T^2/\mu_F^2) 
+ \Bigl. {\cal O}(1/N) \Bigr] \;\;,
\eeq
where 
\beq
\label{c12coefphot}
c_{1,2}^{(q\gamma)} = \frac{1}{2\pi} C_F  \;,
%\eeq
\quad
%\beq
%\label{c11coefphot}
c_{1,1}^{(q\gamma)}(E_T^2/\mu_F^2) = \frac{1}{\pi} C_F
\left( \GE  
+ \frac{3}{4} - \ln \frac{2E_T^2}{\mu_F^2} \;\right) \;,
%\eeq
\quad
%\beq
%\label{c10coefphot}
c_{1,0}^{(q\gamma)}(E_T^2/\mu_F^2) = \frac{1}{\pi}
C_{q\gamma\to \gamma}^{(1)}(2E_T^2/\mu_F^2) \;.
\eeq
This result agrees with the large-$N$ limit of the NLO analytic
expressions computed in Ref.~\cite{aurphot}.

At NNLO we predict
\beeq
{\hat \sigma}_{q\gamma\to \gamma, \,N}^{(2)}(E_T^2, \mu^2, \mu_F^2, \mu_f^2) &=&
{\hat \sigma}_{q\gamma\to q\gamma, \,N}^{(0)} \;
\Bigl[ c_{2,4}^{(q\gamma)} \;\ln^4 N + c_{2,3}^{(q\gamma)}(E_T^2/\mu_F^2) \ln^3 N
\Bigr. \nonumber \\
\label{sig2Nphot} 
&+& \Bigl. c_{2,2}^{(q\gamma)}(E_T^2/\mu_F^2,E_T^2/\mu^2) \ln^2 N
+ {\cal O}(\ln N) \Bigr] \;\;,
\eeeq
where
\beq
\label{c24coefphot}
c_{2,4}^{(q\gamma)} = \frac{1}{2} \;\left[ c_{1,2}^{(q\gamma)} \right]^2 
= \frac{1}{8 \pi^2} C_F\;\;,
\eeq
\beq
\label{c23coefphot}
c_{2,3}^{(q\gamma)}(E_T^2/\mu_F^2) = 
c_{1,2}^{(q\gamma)} \, c_{1,1}^{(q\gamma)}(E_T^2/\mu_F^2)
+ \frac{1}{6\pi} C_F b_0   
= \frac{1}{2\pi^2} C_F \left[ C_F \left( \GE + \frac{3}{4} - 
\ln \frac{2 E_T^2}{\mu_F^2} \right) + \frac{\pi}{3} b_0 \right]
\;\;,
\eeq
%\beeq
%&&c_{2,2}^{(q\gamma)}(E_T^2/\mu_F^2,E_T^2/\mu^2) =
%\frac{1}{2} \;\left[ c_{1,1}^{(q\gamma)}(E_T^2/\mu_F^2) \right]^2
%+ c_{1,2}^{(q\gamma)} \;c_{1,0}^{(q\gamma)}(E_T^2/\mu_F^2)
%\nonumber \\
%&&\;\; \;\;\;\;+ \frac{1}{2\pi} C_F b_0 \left[ \GE   
%+ \frac{3}{4} + \frac{K}{2 \pi b_0} - \ln \frac{2 E_T^2}{\mu^2} 
%\right] \;\;, 
%\eeeq
\beeq
c_{2,2}^{(q\gamma)}\!\left(\frac{E_T^2}{\mu_F^2},\frac{E_T^2}{\mu^2}\right) &=&
\frac{1}{2} \;\left[ c_{1,1}^{(q\gamma)}(E_T^2/\mu_F^2) \right]^2
+ c_{1,2}^{(q\gamma)} \;c_{1,0}^{(q\gamma)}(E_T^2/\mu_F^2)
+ \frac{C_F b_0}{2\pi} \left( \GE   
+ \frac{3}{4} + \frac{K}{2 \pi b_0} - \ln \frac{2 E_T^2}{\mu^2} 
\right) \nonumber \\
&=& \frac{C_F}{2\pi^2} \left[ C_{q\gamma\to \gamma}^{(1)}(2E_T^2/\mu_F^2)
+ \Bigl( C_F + \pi b_0 \Bigr) \left( \GE + \frac{3}{4} - 
\ln \frac{2 E_T^2}{\mu^2} \right) + \frac{1}{2} K \right]
\;, 
\eeeq
and the coefficient $K$ is given in Eqs.~(\ref{kcoef}).


\newpage
                                         
\begin{thebibliography}{99}                                            
%\def    \nuke   #1#2#3{{Nucl. Phys.} {\bf B#1}  (19#2) #3}
\def    \sjnp    #1#2#3{{Sov. J. Nucl. Phys.} {\bf #1} (19#2) #3}
\def    \np     #1#2#3{{Nucl. Phys.} {\bf B#1} (19#2) #3}
\def    \prep   #1#2#3{{Phys. Rep.} {\bf #1}  (19#2) #3}   
\def    \pl     #1#2#3{{Phys. Lett.} {\bf B#1} (19#2) #3}
%\def    \plold  #1#2#3{{Phys. Lett.} {\bf #1B} (19#2) #3}
\def    \prl    #1#2#3{{Phys. Rev. Lett.} {\bf #1}  (19#2) #3}
\def    \pr     #1#2#3{{Phys. Rev.} {\bf D#1}  (19#2) #3}
%\def    \prd    #1#2#3{{Phys. Rev.} {\bf D#1}  (19#2) #3}
\def    \zeit   #1#2#3{{Z. Phys.} {\bf C#1}  (19#2) #3}
\def    \cmp    #1#2#3{{Comm. Math. Phys.} {\bf #1}  (19#2) #3}
\def    \ibid   #1#2#3{{\it ibid.} {\bf #1} (19#2) #3}    
\def    \hepph  #1 {{\tt hep-ph/#1}}
\def    \hepex  #1 {{\tt hep-ex/#1}}
\parskip 0pt
\itemsep=0pt
%\small
%
\bibitem{aurenche}
      P.\ Aurenche, R.\ Baier, A.\ Douiri, M.\ Fontannaz and D.\ Schiff,
      \pl{140}{84}{87};\\
      P.\ Aurenche, R.\ Baier, M.\ Fontannaz and D.\ Schiff,
      \np{297}{88}{661}.
\bibitem{baer90}
   % A NEXT-TO-LEADING LOGARITHM CALCULATION OF DIRECT PHOTON PRODUCTION
   H. Baer, J. Ohnemus and J.F. Owens, \pr{42}{90}{61};\\
   P.\ Aurenche, R.\ Baier and M.\ Fontannaz, \pr{42}{90}{1440}.
\bibitem{gordon93}
   % POLARIZED AND UNPOLARIZED PROMPT PHOTON PRODUCTION BEYOND THE LEADING ORDER
    L.E.\ Gordon and W.\ Vogelsang, \pr{48}{93}{3136}. 
\bibitem{berger91}
   E.L. Berger and J. Qiu, \pl{248}{90}{371}, \pr{44}{91}{2002}.
\bibitem{berger}
     E.L.\ Berger, X.\ Guo and J.\ Qiu, \prl{76}{96}{2234},
     \pr{54}{96}{5470}, hep-ph/9708408, published
    in Proc. of the {\it 32nd Rencontres de Moriond: QCD and High-Energy
    Hadronic Interactions}, ed. J. Tran Than Van (Editions Frontieres, Paris,
    1997), p.~267.
\bibitem{AFGKP}
    P.\ Aurenche, M.\ Fontannaz, J.Ph.\ Guillet, A.\ Kotikov and E.\ Pilon, 
    \pr{55}{97}{R1124}.
\bibitem{pilon}
    S.\ Catani, M.\ Fontannaz and E.\ Pilon,  preprint CERN-TH-98-82
    (\hepph{9803475}).    
\bibitem{bailey92}
   B.\ Bailey, J. Ohnemus and J.F.\ Owens, \pr{46}{92}{2018}.
\bibitem{gordon94}
   % POLARIZED AND UNPOLARIZED ISOLATED PROMPT PHOTON PRODUCTION BEYOND THE LO
   L.E.\ Gordon and W.\ Vogelsang, \pr{50}{94}{1901}.     
\bibitem{ACGG}                      
%QCD CORRECTIONS TO PARTON-PARTON SCATTERING PROCESSES.
    F.\ Aversa, P.\ Chiappetta, M.\ Greco and J.Ph. Guillet,
    \np{327}{89}{105}.
\bibitem{aurenche89}
   P.\ Aurenche, R.\ Baier, M.\ Fontannaz, J.F.\ Owens and M.\ Werlen, 
   \pr{39}{89}{3275}.
\bibitem{gluck94}                                     
    % HIGH P(T) PHOTON PRODUCTION AT P ANTI-P COLLIDER.
    M. Gluck, L.E. Gordon, E. Reya and W. Vogelsang, \prl{73}{94}{388}.
\bibitem{huston95}
    % A GLOBAL QCD STUDY OF DIRECT PHOTON PRODUCTION.
    J. Huston et al., \pr{51}{95}{6139}.
\bibitem{vogelsang95}  
    % CONSTRAINTS ON THE PROTON'S GLUON DISTRIBUTION FROM PROMPT PHOTON
    % PRODUCTION.
    W. Vogelsang and A. Vogt, \np{453}{95}{334}.
\bibitem{WA70}
  %  A MEASUREMENT OF DIRECT PHOTON PRODUCTION AT LARGE TRANSVERSE MOMENTUM 
  %  IN PI-P, PI+ P AND P P COLLISIONS AT 300-GEV/C.   
  C. De Marzo et al., NA24 Collaboration,
  \pr{36}{87}{8}; \\
  M. Bonesini et al., WA70 Collaboration,
  \zeit{38}{88}{371};\\
  % DIRECT PHOTON PRODUCTION IN PROTON - ANTI-PROTON INTERACTIONS AT S**(1/2) =
  % 24.3-GEV.
  A. Bernasconi et al., UA6 Collaboration,
  \pl{206}{88}{163}.
\bibitem{E706-93}
   G. Alverson et al., E706 Collaboration, \pr{48}{93}{5}.
\bibitem{ISR}    %ISR
   %  DIRECT PHOTON PRODUCTION AT THE CERN ISR
   A.L.S. Angelis et al., CMOR Collaboration, \np{327}{89}{541};\\
   %  HIGH P(T) DIRECT PHOTON PRODUCTION IN P P COLLISIONS.
   E. Anassontzis, R806 Collaboration, \zeit{13}{82}{277};\\
   T. Akesson et al., R806 Colaboration,                      
   \sjnp{51}{90}{836}.
\bibitem{UA1-UA2}
    %DIRECT PHOTON PRODUCTION AT THE CERN PROTON - ANTI-PROTON COLLIDER
    C. Albajar et al., UA1 Collaboration, \pl{209}{88}{385};\\
    % A MEASUREMENT OF THE DIRECT PHOTON PRODUCTION CROSS-SECTION AT THE CERN 
    %  ANTI-P P COLLIDER.
    J. Alitti et al., UA2 Collaboration, \pl{263}{91}{544}.            
\bibitem{CDF-D0}      
  F. Abe et al., CDF Collaboration, \prl{73}{94}{2662}; \\
  S. Abachi et al., D0 Collaboration, \prl{77}{96}{5011}.
\bibitem{fmnr}
S. Frixione, M.L. Mangano, P. Nason and G. Ridolfi, \np{431}{94}{453}, 
   \ibid{405}{93}{507}, preprint CERN-TH-97-16 (\hepph{9702287}),
to be published in Heavy Flavours II, ed. by A.J. Buras and M. Lindner, World
Scientific. 
\bibitem{baer96}
    %MULTIPLE PARTON EMISSION EFFECTS IN NEXT-TO-LEADING ORDER DIRECT PHOTON
    %PRODUCTION.
    H. Baer, M.H. Reno, \pr{54}{96}{2017}.
\bibitem{vogelsang98}
   W. Vogelsang, private communication.
\bibitem{E706-97}                                         
   L. Apanasevich et al., E706 Collaboration, preprint FERMILAB--Pub--97/351-E
   (\hepex{9711017}).
\bibitem{MRST}
  A.D. Martin, R.G. Roberts, W.J. Stirling and R.S. Thorne,
   preprint DTP/98/10, RAL-TR-98-029 (\hepph{9803445}).
\bibitem{fontannaz}
  M.\ Fontannaz,
  in Proc. of the {\it 32nd Rencontres de Moriond: QCD and High-Energy
    Hadronic Interactions}, ed. J. Tran Than Van (Editions Fronti\`eres, Paris,
    1997), p.~235.
\bibitem{cdfjets}
  F. Abe et al., CDF Collaboration, 
  \prl{77}{96}{438}.
\bibitem{bonciani}
    R. Bonciani, University of Florence Thesis, July 1997 (in Italian).
\bibitem{BCMN}
    R.~Bonciani, S.~Catani, M.L.~Mangano and P.~Nason, preprint CERN-TH-98-31
    (\hepph{9801375}),  to appear in Nucl. Phys. B. 
\bibitem{inprep}                               
    R.~Bonciani, S.~Catani, M.L.~Mangano and P.~Nason, in preparation.
\bibitem{Laenen98}
 E. Laenen, G. Oderda and G. Sterman, \hepph{9806467}.
\bibitem{Sterman}   
  G.\ Sterman, \np{281}{87}{310}.
\bibitem{CT}
  S.\ Catani and L.\ Trentadue, \np{327}{89}{323}.
\bibitem{CT2}  
  S.\ Catani and L.\ Trentadue, \np{353}{91}{183}.
\bibitem{SC}
  See: G.\ Sterman, in Proc. {\it 10th Topical Workshop on Proton-Antiproton
  Collider Physics}, eds. R.\ Raja and J.\ Yoh (AIP Press, New York, 1996),
  p.~608; 
  S.\ Catani, 
%  hep-ph/9709503, published
    in Proc. of the {\it 32nd Rencontres de Moriond: QCD and High-Energy
    Hadronic Interactions}, ed. J. Tran Than Van (Editions Fronti\`eres, Paris,
    1997), p.~331; and references therein.
\bibitem{BCM}
  A.\ Bassetto, M.\ Ciafaloni and G.\ Marchesini, \prep{100}{83}{201}; 
  Yu.L.\ Dokshitzer, V.A.\ Khoze, A.H.\ Mueller and  S.I.\ Troyan,
  {\it Basics of Perturbative QCD} (Editions Fronti\`eres, Gif-sur-Yvette,
  1991). 
\bibitem{kidon}
%    N.~Kidonakis, PhD thesis (SUNY, Stony Brook), 
%    \hepph{9606474};\\                            
    N.\ Kidonakis and G. Sterman, \pl{387}{96}{867}, 
%    preprint EDINBURGH-97-3 (hep-ph/9705234).
    \np{505}{97}{321}.
\bibitem{jetnlo}
    N.\ Kidonakis, G.\ Oderda and G.\ Sterman, preprint EDINBURGH-97-22 
    (\hepph{9801268}).
\bibitem{kos}
    N. Kidonakis, G. Oderda and G. Sterman, preprint ITP-SB-98-23
    (\hepph{9803241}).
\bibitem{KT}     
   J.~Kodaira and L.~Trentadue, \pl{112}{82}{66}.  
\bibitem{CdET} 
   S.~Catani, E.~d'Emilio and L.~Trentadue, \pl{211}{88}{335}. 
\bibitem{CMNT2}  
   S. Catani, M.L. Mangano, P. Nason and L. Trentadue, \pl{351}{96}{555};
   \np{478}{96}{273}.
%  (hep-ph/9604351).
\bibitem{contogouris}
      A.P.\ Contogouris, N.\ Mebarki and S.\ Papadopoulos, Int. J. Mod. Phys.
      {\bf A5} (1990) 1951; A.P.\ Contogouris and S.\ Papadopoulos, Mod. Phys. 
      Lett. {\bf A5} (1990) 901.
\bibitem{CTTW}   
  S. Catani, L. Trentadue, G. Turnock and B.R. Webber,
  \np{407}{93}{3}.
\bibitem{CMW}              
  S. Catani, G. Marchesini and B.R. Webber, \np{349}{91}{635}.
\bibitem{Spira}
 M. Kr\"amer, E. Laenen and M. Spira, \np{511}{98}{523}.
% preprint CERN-TH/96-231, \hepph{9611272}.
\bibitem{Sterman96}
  H.\ Contopanagos, E.\ Laenen and G.\ Sterman, \np{484}{97}{303}.
%   hep-ph/9604313.
\bibitem{thrust}
   S. Catani, G. Turnock, B.R. Webber and L. Trentadue, \pl{263}{91}{491}
\bibitem{aurphot}
      P.\ Aurenche, A.\ Douiri, R.\ Baier, M.\ Fontannaz and D.\ Schiff,
      \zeit{24}{84}{309}.
\end{thebibliography}
\end{document}